\PassOptionsToPackage{dvipsnames}{xcolor}
\documentclass[fleqn,12pt,a4paper]{article}
\usepackage[a4paper, top=1in, 
left=1in, right=1in, bottom=1in]{geometry}

\usepackage[colorinlistoftodos]{todonotes}
\usepackage{multicol}
\usepackage{theorem}
\usepackage{amsmath}
\usepackage{amssymb}
\usepackage{amsopn}
\usepackage{latexsym}
\usepackage{setspace}
\usepackage{dsfont}
\usepackage{datetime}
\usepackage[allnumber,warning,easyold,math]{easyeqn}
\usepackage{xcolor}
\usepackage[12pt]{moresize}
\usepackage[bookmarks=true,bookmarksnumbered=true,
pdfpagemode=None,pdfstartview=FitH,hidelinks]{hyperref}
\usepackage{fix-cm}
\usepackage[calcwidth]{titlesec}
\usepackage{graphicx}
\usepackage{sgame}
\usepackage[utf8]{inputenc}
\usepackage[english]{babel}
\usepackage{braket}
\usepackage{upgreek}
\usepackage[footnotesize]{caption}
\usepackage{tikz}
\usetikzlibrary{arrows.meta, positioning, shadows, shapes.symbols}
\usepackage[font=scriptsize]{subfig}
\usepackage{multirow}
\usepackage[normalem]{ulem}
\usepackage[style=authoryear,maxnames=99,backend=bibtex,natbib=true,maxcitenames=5]{biblatex}

\theoremstyle{break}
\theorembodyfont{\itshape}

\newtheorem{thm}{Theorem}
\newtheorem{prp}[thm]{Proposition}

\newtheorem{obs}{Observation}

\theorembodyfont{\upshape}

\newcommand{\sqbox}{\hfill$\blacksquare$}

\newcommand{\R}{\mathbb{R}}
\def \cm{{\mathcal M}}

\bibliography{DCLOGIT.bib}

\begin{document}

\title{\Large \textbf{Decision Conflict, Power Logit, \linebreak 
and the Deferral Outside Option}}
	
\author{\vspace{-5pt} Georgios Gerasimou\thanks{Email address:
\href{mailto: Georgios.Gerasimou@glasgow.ac.uk}{
	\color{blue}Georgios.Gerasimou@glasgow.ac.uk}}
\thanks{This paper originates in---and supersedes---work 
originally circulated  
as \textit{``The decision-conflict and multicriteria logit''}
that was originally posted online in May 2020.
The author has benefited from discussing 
this work with many scholars, and is particularly thankful to 
Miguel Ballester, J\"{o}rg Stoye, 
Dimitris Christelis, Irina Merkurieva  
and Levent \"{U}lk\"{u} for their comments;  
to Sudeep Bhatia and Tim Mullett for sharing 
the data that are analysed in Section 5; 
the Economics Department at Oxford University 
for their hospitality in October 2019; and audiences 
at Seminars in Economic Theory (2021), 
Heriott-Watt (2021), SAET 2021, BRIC 2022 (Prague), 
Glasgow, Durham Economic Theory Conference 2023, 
EEA-ESEM 2023 (Barcelona) and Oligo 2024 (Crete).}\\ 
\vspace{10pt}
\footnotesize University of Glasgow\ \\  
\footnotesize \today \vspace{-20pt}}
\date{}
\maketitle
		
\begin{abstract}
Decision makers often opt for the deferral outside option 
when they find it difficult to make an active choice.  
Contrary to existing logit models with an outside option 
where the latter is assigned a fixed value exogenously, 
this paper introduces and studies a class of logit models 
where that option's value is menu-dependent, 
may be determined endogenously, 
and could be interpreted as proxying the vary\nolinebreak ing degree of 
decision difficulty at different menus.
We focus on the \textit{power logit} special class of these models. 
We show that these explain some observed choice-deferral 
effects that are caused by hard decisions, including non-monotonic
``roller-coaster'' choice-overload phenomena that are regulated 
by the presence or absence of a clearly dominant feasible alternative.
We illustrate the usability, novel insights and explanatory gains 
of the proposed framework for duopolistic modelling and 
empirical discrete choice analysis.
\end{abstract}
		
\vfill
			
\thispagestyle{empty}	

\pagebreak
	
\setcounter{page}{1}

\normalsize

\section{Introduction}

It is a well-established fact that people often 
``choose not to choose''
when they find it hard to compare 
the active-choice alternatives available to them, 
even when all these alternatives are individually considered ``good enough'' 
and are paid attention to. 
Real-world examples of such behavior include: 
(i) employees who operated within an ``active decision'' pension-savings 
environment and did not sign up for one of the plans that were available 
to them within, say, a day, week or month of first notice, 
possibly even opting for indefinite non-enrolment;\footnote{Such behavior 
is documented in \textcite{carroll-choi-laibson-madrian-metrick}, for example.} 
(ii) patients who, instead of choosing \textit{``immediately''} one of 
the active treatments that were recommended to them against a medical condition, 
delayed making such a choice---often at a health cost---due 
to \textit{``facing a treatment dilemma''};\footnote{See 
\textcite[p. 78]{knops-goossens-ubbink-legemate-stalpers-bossuyt13}. See also
\citet{oconnor95,DecConfScale19} for additional references and overview
of the use of a \textit{``decisional conflict scale''} in medical 
decision making that was developed \textit{``to measure a person's 
perceptions of their uncertainty in making a choice about 
health care options, the modifiable factors contributing 
to uncertainty, and the quality of the decision made''}.} 
(iii) doctors who were willing to prescribe the single available drug 
to treat a medical condition but were not prepared to prescribe anything 
when they had to decide from the expanded set that contained one more drug, 
because \textit{``the difficulty in deciding between the two medications 
led some physicians to recommend not starting either''} 
\parencite{redelmeier-shafir95}.
\footnote{Other works that find evidence associating decision difficulty
with such choice paralysis in different environments include 
\citet{tversky-shafir92,dhar97,dhar-simonson03,danan&ziegelmeyer06,
bhatia-mullett16,CCGT22}.}
Even though in many such cases people  
do end up making an active choice \textit{eventually}, 
the following question arises for the decision analyst:
\textit{Should the observed individual's initial avoidance/delay to 
make such a choice be ignored as non-relevant?} 
Put more concretely, should the decision of a 
medical doctor who eventually prescribes one of the two possible 
medications after several reminders be thought of as 
similar to their decision to prescribe that medication without delay
in a different situation? 
In this paper we regard the decision to avoid/delay making an active choice 
that is caused by such context-specific ``decision conflict'' as providing meaningful 
information about both the individual and the alternatives in question, regardless of 
which active choice---if any---was eventually made at that menu at a later time.

In their influential monograph, \citet[p. 46]{janis-mann77} defined 
\textit{``decision conflicts''} as the \textit{``simultaneous opposing tendencies
to accept and reject a given course of action'' and identified 
``hesitation, vacillation, \textnormal{[and]} feelings of uncertainty''}
to be among their most prominent symptoms  
\textit{``whenever the decision comes within 
the focus of attention''}.\footnote{\citet{pochonetal08} is a targeted 
study in the neuroscience literature on the brain regions that are activated 
when subjects face decision conflict.}
Motivated by the relevance of hesitation-driven 
opt-out decisions for understanding 
preferences and explaining behavior, our specific goal in this paper is to model 
choice in the presence of a choice-avoidance/deferral outside option within 
a stochastic choice framework in ways that deviate as little 
as possible from existing well-understood modelling practices and, 
at the same time, make predictions that are in line with some findings
from the empirical/experimental literature and evade existing models. 
We pursue this by extending in disciplined ways the foundational 
\textcite{luce59}/logit model and its econometric specification 
pioneered by \textcite{mcfadden73}. Specifically, we propose 
the class of \textit{decision-conflict logit}
models which, in their most general form, are a straightforward but so 
far unexplored extension of the logit with an outside option that 
assign a menu-dependent value to that option while retaining 
the menu-invariance assumption on all active-choice alternatives. 
The relative value of the outside option at a menu in turn 
determines the probability of avoiding/deferring choice and can be 
interpreted as proxying decision difficulty. 

We show that, despite its simplicity, this general model can be  
the starting point for richly structured special cases. 
In particular, we introduce and focus on the broad class of 
\textit{power logit} models 
that are examples of of such cases where decision difficulty 
depends in intuitive ways on the logit values of all active-choice alternatives. 
In these models, decision difficulty could be thought 
of as driven by the agent's noisy resampling of the menu's elements.
In the \textit{quadratic logit} special case of this class
of models, for example, such resampling takes the form of the choice probability 
of a market alternative emerging as the product of 
\textit{two} logit probabilities according to a \textit{single}
 value function/criterion. 
Intuitively, the agent is more likely to choose an active-choice alternative 
if and only if its value realizations according to this criterion are much larger 
than those of everything else feasible across \textit{both} rounds of sampling. 
Conversely, the agent is more likely to avoid/defer choice when no 
alternative achieves such unanimous clear dominance. 
This model could therefore be thought of as capturing a hesitant 
decision maker who behaves as if they used an objective criterion 
to compare alternatives 
(e.g. sum or multiply each option's values across all relevant attributes) 
but is aware that their subjective evaluation according to this 
objective criterion may be imperfect, possibly due to cognitive limitations, 
thereby leading them to performing this task twice. 
This model and its power-logit generalization 
appear to be the first to provide a theory where the no-choice outside option is feasible 
and has an endogenously determined, menu-dependent value.

We further show that these structured models explain the following 
empirical phenomena that various studies in cognitive and consumer 
psychology have documented about decisions that allow agents to 
avoid/delay making an active choice:

\begin{enumerate}
	\item[(i)] As alternatives become 
	more similar in their overall appeal, decision difficulty 
	and the probability of choice delay are increased.
	We will refer to this as the \textit{``similarity-driven deferral effect''}.

\vspace{-5pt}

	\item[(ii)] The dominance-driven non-monotonic relation 
	between menu expansion and the tendency to opt out, 
	which we refer to as the \textit{``roller-coaster'' choice-overload effect}. 
	This has implications for the interpretation and policy 
	responses to so-called ``too-much-choice'' phenomena.

\vspace{-5pt}
	
	\item[(iii)] \textit{``Relative-desirability''} effects, 
	whereby holding constant the total value in a menu 
	while increasing the value differences between the menu's 
	alternatives increases the probability of an active choice.
\end{enumerate}

We illustrate the applicability of our analysis both in theoretical 
and empirical settings. In our theoretical application 
we analyse a duopolistic market 
where firms compete simultaneously in price and quality 
under the common-knowledge assumption that consumer demand is 
determined by the power-logit model where a product's value 
is defined, intuitively, by its quality/price ratio. 
We derive simple and economically interpretable intuitive closed-form 
solutions for all equilibrium variables in the model: price, quality,
profits and a notion of consumer welfare that appears suitable 
in environments where consumers opt out due to indecisiveness or overload.
A key feature of the (symmetric) equilibrium is that, as the power 
parameter capturing consumers' decision difficulty
increases, firms increase their products' quality/price ratio 
and see their profits decreased, both because of the reduced
profit margins and because of the lower share of consumers 
who buy any product. Intuitively, this is driven by each firm 
increasing its quality/price ratio in an effort to reduce 
the consumer's decision difficulty and mitigate the 
risk of losing them to the rival firm \textit{or} of 
driving them out of the market altogether.

In our empirical application we first show 
how the classic assumptions and argument that underpin 
the discrete-choice formulation of the logit \textit{without} 
an outside option \parencite{mcfadden73} must be modified and extended 
in order for both the quadratic logit and the more general power logit 
models to admit a similar discrete-choice formulation and be taken to the data
for maximum-likelihood estimation of their respective parameters.
We then show the potential fruitfulness of such analyses by 
estimating both the quadratic and power logit models on the 
deferral-permitting discrete-choice data with film decisions 
from the survey experiment of \textcite{bhatia-mullett16}, 
using the participants' subjective ratings of the different films 
as the explanatory variable. To assess the added value of the hereby proposed models
on these data, we use standard criteria to evaluate their goodness of fit and compare
them to those of baseline logit models with a fixed and inferior or a 
random outside option. 
Our analysis suggests that both the power and quadratic logit often 
perform better compared to either version of the baseline logit under 
these performance criteria, particularly in those situations where theory 
suggests they would do so. 
Hence, the proposed power-logit framework could be considered in the analysis of 
similar datasets whenever the researcher suspects that the observed 
opting-out/deferring behavior might be due to decision difficulty rather 
than the relative unattractiveness of the available active-choice alternatives.

\section{Theory}

\subsection{Logit with A Menu-Dependent Outside Option} 

Let $X$ be the grand choice set of finitely many 
\textit{active-choice} alternatives with generic elements $a,b\in X$. 
Let $\mathcal{M}:=\{A: \emptyset\neq A\subseteq X\}$ be the collection 
of all \textit{menus} of such alternatives, and $\mathcal{B}$ its 
sub-collection that comprises all binary menus. 
The \textit{outside option} is denoted by $o\not\in X$.
We clarify that this is not a \textit{status quo} option 
(e.g. a tenant's current rental agreement) which can, 
in principle, be compared to the other feasible alternatives 
(e.g. other housing options) on the same or a similar set 
of relevant attributes. 
Instead, this option is devoid of attributes and its value to the 
decision maker 
is unobservable to the analyst.\footnote{\citet[p.53]{hensher-rose-greene15}
, for example, describe this distinction as follows: 
\textit{``At this point, it is worthwhile considering choice situations 
in which there exists the possibility to `choose not to choose', 
or to remain with some status quo alternative. 
Many choice situations present decision makers with examples of both types
of alternatives. 
For example, a person can elect to stay at home and not see
a movie if three potential movie alternatives showing 
at a local cinema at some preferred time do not appeal to them. 
Likewise, a decision maker facing the expiration of their rental 
agreement may elect to simply renew their current rental 
contract or move apartments, hence signing a new lease. 
In the case of a \textbf{no choice} alternative, 
the alternative labelled \textnormal{`none'} will be devoid of any attribute
levels (e.g., there is no movie ticket price, no time spent at the cinema, etc., 
associated with going to the movies). The absence of attributes, however, does not
mean that the decision maker is indifferent to that alternative. In the movie ex-
ample, if the three movies on offer are romantic comedies, then staying at home
and not attending any of them might be the most preferred option.''}
A formal distinction in the treatment of status-quo and choice-deferral 
outside options in a deterministic choice-theoretic framework 
is provided in \citet{gerasimou16a}.}
A \textit{random free-choice model} on $X$ is a function 
$\rho:X\times \mathcal{M}\rightarrow\R_{+}$ such that 
$\rho(a,A)\in [0,1]$ for all $A\in\mathcal{M}$ and all $a\in A$; 
$\rho(a,A)=0$ for all $A\in\cm$ and all $a\not\in A$; 
and $\sum_{a\in A} \rho(a,A)\leq 1$, 
where $\rho(o,A):=1-\sum_{a\in A} \rho(a,A)\leq 1$ is the 
probability of choosing the---always feasible---outside option at menu $A$. 
To simplify notation, for $A,B\in\cm$ with $B\subseteq A$ we write
$\rho(B,A):=\sum_{b\in B}\rho(b,A)$. 

We start by introducing the \textit{logit with a general outside option} as
the model that comprises value functions $u:X\rightarrow\R_{++}$ and 
$D:\mathcal{M}\rightarrow\R_{+}$ such that, for every menu $A\in\cm$ and 
active-choice alternative $a\in A$,
\begin{eqnarray}\label{first}
	\rho(a,A) 
	& = 
	&  \displaystyle\frac{u(a)}{\displaystyle\sum_{b\in A}u(b)+D(A)},
\end{eqnarray} 
where the pair $(u,D)$ is unique up to a common positive linear transformation.
In this model, $u$ captures the menu-independent values of 
active-choice alternatives and $D(A)$ the menu-dependent value 
of the outside option. 
Like the baseline Luce model [see \eqref{logit-without} below], 
all active-choice alternatives in \eqref{first}
are assigned menu-independent values that determine their relative likelihood 
of being chosen. Unlike the baseline model, 
where this property also extends to the outside option, 
here the probability of making an active choice in the first place 
(equivalently, of avoiding/deferring this decision) 
is determined by the menu-dependent value of $D$.

Two properties characterize the class of models that can be represented 
in this way:\\
	
\noindent\textbf{A1 (Positivity)}.\\ 
\textit{For all $A\in\cm$ and all $a\in A$: $\rho(a,A)>0$.}\\

\noindent\textbf{A2 (The Active-Choice Luce Axiom)}.\\ 
\textit{For all $A,B\in\cm$ and all $a,b\in A\cap B$:}
\begin{eqnarray}
	\nonumber \frac{\rho(a,A)}{\rho(b,A)} 
	& = 
	& \frac{\rho(a,B)}{\rho(b,B)}
\end{eqnarray}

\noindent A1 is standard and allows for a crisp illustration of the main 
ideas that we put forward in this paper. 
A2 imposes the familiar kind of IIA-consistency only in the odds of pairs 
of active-choice alternatives, while allowing odds that involve 
such an alternative and the outside option to deviate from it. 
That is, $\frac{\rho(o,A)}{\rho(b,A)}\neq \frac{\rho(o,B)}{\rho(b,B)}$ 
is allowed by A2.

\begin{prp}
	\label{characterization}
	$\rho$ is a logit with a general outside option 
	if and only if it satisfies A1-A2.
\end{prp}

Indeed, adapting the arguments in \textcite{luce59} yields an 
equivalence between A1-A2 and the existence of a function 
$u:X\rightarrow\R_{++}$ such that, for every $A\in\cm$ and $a\in A$,
\begin{eqnarray}
	\label{luce-0} \rho(a,A) 
	& = 
	& \big(1-\rho(o,A)\bigr)\cdot \frac{u(a)}{\displaystyle\sum_{b\in A}u(b)},
\end{eqnarray}
where
\begin{eqnarray}
	\label{u-definition} u(a) 
	& := 
	& \alpha\cdot\frac{\rho(a,X)}{\rho(z,X)}
\end{eqnarray}
for arbitrary and fixed $\alpha>0$ and $z\in X$. 
It follows then that for every $A\in\cm$ there is a unique $D(A)\geq 0$ 
that makes \eqref{first} true, with
\begin{eqnarray}\label{D-definition} 
	D(A) & \equiv & \frac{\rho(o,A)}{1-\rho(o,A)}\cdot \sum_{b\in A} u(b).
\end{eqnarray}
Finally, it is immediate that $(u,D)$ and $(u',D')$ represent the same 
$\rho$ if and only if $u=\alpha u'$ and $D=\alpha D'$ for some $\alpha>0$.

We now compare \eqref{first} to the baseline logit 
\textit{with} an outside option 
\parencite{anderson_etal,hensher-rose-greene15} and 
\textit{without} such an option. 
We recall that a random choice model 
$\rho$ on $X\cup\{o\}$ admits the former representation 
if there is a function $u:X\cup\{o\}\rightarrow\R_{++}$ such that, 
for all $A\in\cm$ and $a\in A$,
\begin{eqnarray}\label{logit-with} 
	\rho(a,A) & = & \frac{u(a)}{\displaystyle\sum_{b\in A}u(b)+u(o)}.
\end{eqnarray}
On the other hand, $\rho$ admits a logit representation 
\textit{without} an outside option if there exists some 
$u:X\rightarrow\R_{++}$ such that
\begin{eqnarray}\label{logit-without} 
	\rho(a,A) & = & \frac{u(a)}{\displaystyle\sum_{b\in A}u(b)},
\end{eqnarray}
The latter obviously implies $\sum_{a\in A}\rho(a,A)=1$ for all 
$A\in\cm$, so that the opportunity to defer is either infeasible 
in this model or feasible but never acted upon.
Thus, \eqref{first} includes \eqref{logit-without} as a special case 
when $o\not\in X$. In addition, \eqref{first} 
extends \eqref{logit-with} but without nesting it unless A1-A2 and $\rho$
operate on the enriched domain $X\cup \{o\}$.

\subsection{Decision-Conflict Logit and Power Logit}

We define the \textit{power logit} model by the existence 
of a menu-independent \textit{stimulus intensity} value function 
$\widehat{u}:X\rightarrow\R_{++}$ and a parameter $p\geq 1$ such that, 
for every menu $A$ and alternative $a$ in $A$,
\begin{eqnarray}
	\label{power} \rho(a,A) 
	& = 
	& \left(\frac{\widehat{u}(a)}{\sum\limits_{b\in A} \widehat{u}(b)}\right)^p
\end{eqnarray}
Clearly, this model predicts $\rho(o,A)>0$ at every menu $A$ if and only if $p>1$,
and reduces to \eqref{logit-without} at $p=1$. 

The agent portrayed in \eqref{power} could be thought of as behaving according 
to the standard logit with a single valuation criterion but, possibly aware 
of their decision difficulty, also as if they sampled all alternatives 
\textit{more than once} before making a decision. 
For example, in the \textit{quadratic logit} case of special interest 
where $p=2$, the agent might be thought of as 
sampling the same menu \textit{twice}. 
Because the resulting value realizations generally differ 
across these two rounds of sampling due to the postulated randomness, 
this individual would be more likely to choose an active-choice 
alternative if its perceived \textit{signal/stimulus intensity} 
from \textit{both} inspections, captured by the two value realizations 
of $\widehat{u}$, is relatively high, and as being more likely to 
avoid/defer choice when this is not true for any such alternative. 
When deciding which insurance plan to buy, for example, 
an agent whose behavior is approximated by the quadratic logit may review the 
top-rated plans from a service comparison website in the morning, 
receive some value stimuli/signals from each of them, and then go back and 
repeat this process in the evening. Assuming that the two sampling rounds 
are independent (admittedly, a demanding assumption), an insurance plan 
is more likely to be chosen at the end of this two-stage process if its 
relative stimulus/signal intensity is sufficiently high to make the product 
stand out despite the agent's hesitation. 

The intuition in the more general case where $p\neq 2$ in \eqref{power} 
is analogous and admits a probabilistic explanation. Specifically, 
if the analyst a priori restricts $p$ to lie between 1 and 2, 
then $p-1$ might be interpreted as the (exogenous) 
\textit{probability} that the agent will engage in two rounds of sampling, 
equalling 1 in the limit where the quadratic logit decision process
emerges with certainty. 
Similarly, if $p$ is assumed to lie between 2 and 3, 
then $p-2$ could be thought of as the probability that the agent
will perform three rounds of sampling, conditional on the analyst 
expecting them to do at least two.
More generally, the power parameter $p$ in this model could be viewed 
as reflecting the agent's propensity to engage 
in possibly multiple rounds of sampling.

That this model is a logit with a general outside option
may not be obvious at first glance but quickly becomes 
so upon noticing that one can write

\begin{eqnarray}
	\label{power1} u(a) 
	& := 
	& \widehat{u}(a)^p\\
	\label{power2} D(A) 
	& := 
	& \left(\sum_{b\in A}\widehat{u}(b)\right)^p-\sum_{b\in A}\widehat{u}(b)^p
\end{eqnarray}

\noindent
At $p=2$ these expressions admit the simpler and more easily interpretable form
\begin{eqnarray}
	\label{auxiliary1} u(a) 
	& := 
	& \widehat{u}(a)^2,\\
	\nonumber D(A) 
	& := 
	& \left(\sum_{b\in A}\widehat{u}(b)\right)^2-\sum_{b\in A}\widehat{u}(b)^2 \\
	\nonumber 
	& = 
	& 2\sum_{\substack{a,b\in A,\\ a\neq b}}\widehat{u}(a)\widehat{u}(b)\\
	\label{auxiliary3} & = & \sum_{a,b\in A}D(\{a,b\}),
\end{eqnarray}
where the last step makes use of the notational convention 
\begin{eqnarray*}
	D(\{a,a\}) & \equiv & D(\{a\})
\end{eqnarray*}
This clarifies that the quadratic logit 
$\rho\sim(\widehat{u})^2$ is an \textit{additive} $(u,D)$ 
model in the sense that the value of the outside option at every 
menu depends additively on the value of that option at each 
of its binary submenus. It also clarifies that the latter value 
takes a symmetric Cobb-Douglas form with respect to $\widehat{u}$.
We will return to additivity later in this section but note 
here that the quadratic case where $p=2$ is the only 
one where the $(u,D)$ representation of \eqref{power} has this property.

We proceed by noting the following direct 
implication of the power-logit model:\\

\noindent \textbf{A3 (Desirability \& Complexity)}\\ 
\textit{For all $A\in\cm$: $\rho(A,A)=1$ $\Longleftrightarrow$ $|A|=1$.}\\

\noindent To motivate the intuition behind (and the label assigned to) 
A3  we first recall that, as was clarified early on, 
our aim here is to model decision difficulty that is 
rooted in a fully attentive individual's potential inability to 
make some preference comparisons between otherwise desirable options. 
If a \textit{single} such option was feasible to such an individual, 
therefore, one might expect that person to immediately choose 
that one option. If on the other hand there are at least two available 
options and the individual is not forced to make a choice immediately, 
then the experimental/empirical evidence suggests that there is at 
least \textit{some} probability that this person's attempt to find 
a most preferred option and choose that option will not be fruitful 
reasonably quickly. To the extent that this is so, a legitimate approach 
from the analyst's perspective would be to portray that decision maker 
as \textit{deferring} choice with positive probability whenever 
at least one non-trivial comparison is required.

Imagine, for example, a patient like those reported on in 
\textcite{knops-goossens-ubbink-legemate-stalpers-bossuyt13} who has been 
diagnosed with a life-threatening disease. 
Suppose that their doctor informs them that there is only one available 
treatment that can cure this disease, and asks whether they would like 
to sign up for this treatment. One would expect the patient to sign up 
immediately because there would be no benefit from delaying their only 
chance for a cure. Now suppose instead that the doctor tells the patient 
that there are two possible treatments: one with high efficacy but 
severe side effects, and another with milder side effects but lower 
cure rates. Even though either one of these treatments would 
have been chosen immediately if it was the only feasible one
(see \cite{tversky-shafir92,redelmeier-shafir95,dhar97,CCGT22}, for example),
here one might expect 
the patient to delay making such an active choice, perhaps until 
they think about the conflicting pros and cons and then ultimately 
determine which treatment would be best for them. 
Situations of this kind are compatible with and, in fact, motivate 
our modelling framework in this paper.

In the spirit of these examples, A3 postulates that an active choice 
is made with certainty \textit{only at} singleton menus and, as such, 
formalises the behavioral mechanisms outlined above. 
Of course, one can easily think of situations where this axiom is 
descriptively invalid. Yet for analytical purposes it is a useful 
property because it allows for completely isolating the 
decision-difficulty channel to deferrals from other potential 
channels such as undesirability of the available alternatives 
or limited attention, which have quite distinct behavioral origins.

The following statement readily follows from the preceding
analysis.

\begin{obs}
	\label{dcl}
	$\rho$ satisfies A1--A3 if and only if 
	it is a $(u,D)$-model with the property that
	\begin{eqnarray}
		D(A) = 0 & \Longleftrightarrow & |A|=1
	\end{eqnarray}
\end{obs}

\noindent We will refer to this special class of 
generalized logit models with a context-dependent outside 
option as the class of \textit{decision-conflict logit} models, 
and to the menu function $D$ that captures the varying appeal 
of opting out at different menus as the \textit{decision cost} 
or \textit{decision complexity} function. 
Justifying such a name for the function $D$ given the requirement 
that it be zero-valued only at singletons may benefit from some 
additional explanation that supplements the preceding discussion. 
When the decision environment is such that avoidance/deferral 
is caused solely by decision difficulty instead of other factors 
(e.g. none of the active-choice alternatives is good enough, 
or none is considered due to limited-attention constraints), 
our decision maker is portrayed as not having any problem deciding 
between deferring or choosing the only available active-choice option: 
they do the latter. By contrast, the decision between deferring 
or choosing from two or more such options is \textit{at least} 
somewhat costly because of the effort that is necessary 
to make the relevant preference comparisons. 

We will refer to both a decision-conflict logit 
$\rho=(u,D)$ and $D$ as \textit{monotonic} if
\begin{eqnarray}\label{monotonic}
	A \supset B & \Longrightarrow & D(A) \geq D(B).
\end{eqnarray}
If $D(A)>D(B)$ is always true when $A\supset B$, 
then $D$ and $\rho=(u,D)$ will be called \textit{strictly} monotonic. 
In line with our intended interpretation of $D$ as a complexity/cost function, 
the total number of pairs of distinct alternatives increases as a menu expands, 
hence so does the expected number of comparisons between alternatives 
that a fully-attentive individual needs to make. 
In expectation, therefore, decision difficulty also goes up in absolute 
terms when more alternatives are added to a menu. 
Importantly, however, this \textit{does not} imply that deferring always 
becomes more likely once a menu is expanded when $D$ is monotonic 
(we will return to this point soon). 
But Monotonicity does have a familiar general implication 
for \textit{active}-choice alternatives, which in the standard 
random forced-choice environments was originally stated 
in \textcite{block-marschak60}:

\begin{obs}\label{acr}
	If $\rho$ is a monotonic decision-conflict logit, 
	then $a\in B\subset A$ implies $\rho(a,B)>\rho(a,A)$.
\end{obs}

\noindent
In particular, monotonic models satisfy what we will refer to 
as \textit{active-choice regularity}, whereby the probability 
of such alternatives cannot increase when more options are 
added to a menu. Crucially, however, as we discuss and 
illustrate by example later, this property does not 
hold for the outside option. 

When it comes to using a decision-conflict logit in suitable applications, 
the analyst must first decide whether to employ a special case where
function $D$ is set exogenously or one where it is determined 
endogenously. In the first case the choice might be 
dictated by the analyst's a priori assessment of the specific 
environment in question and could include, for example, 
defining $D$ as the menu-cardinality function 
\parencite{iyengar-lepper00,iyengaretal04} 
or, if the alternatives have clearly identifiable attributes, 
some measure of similarity in attribute space 
\parencite{spektor-gluth-fontanesi-rieskamp19}. 
The analyst's choice in the second
case might instead be dictated by an agnosticism towards what is 
the most appropriate functional form for $D$, and by resorting instead
to a general decision process where $D$ is a function of the feasible options'
$u$-values. The power-logit class of models is clearly of this kind.

\begin{figure}[!htbp]
	\centering
	\caption{\centering The loci of binary choice probability distributions that are 
		compatible \newline with the power logit under different values of $p$.}
		\includegraphics[width=0.65\textwidth]{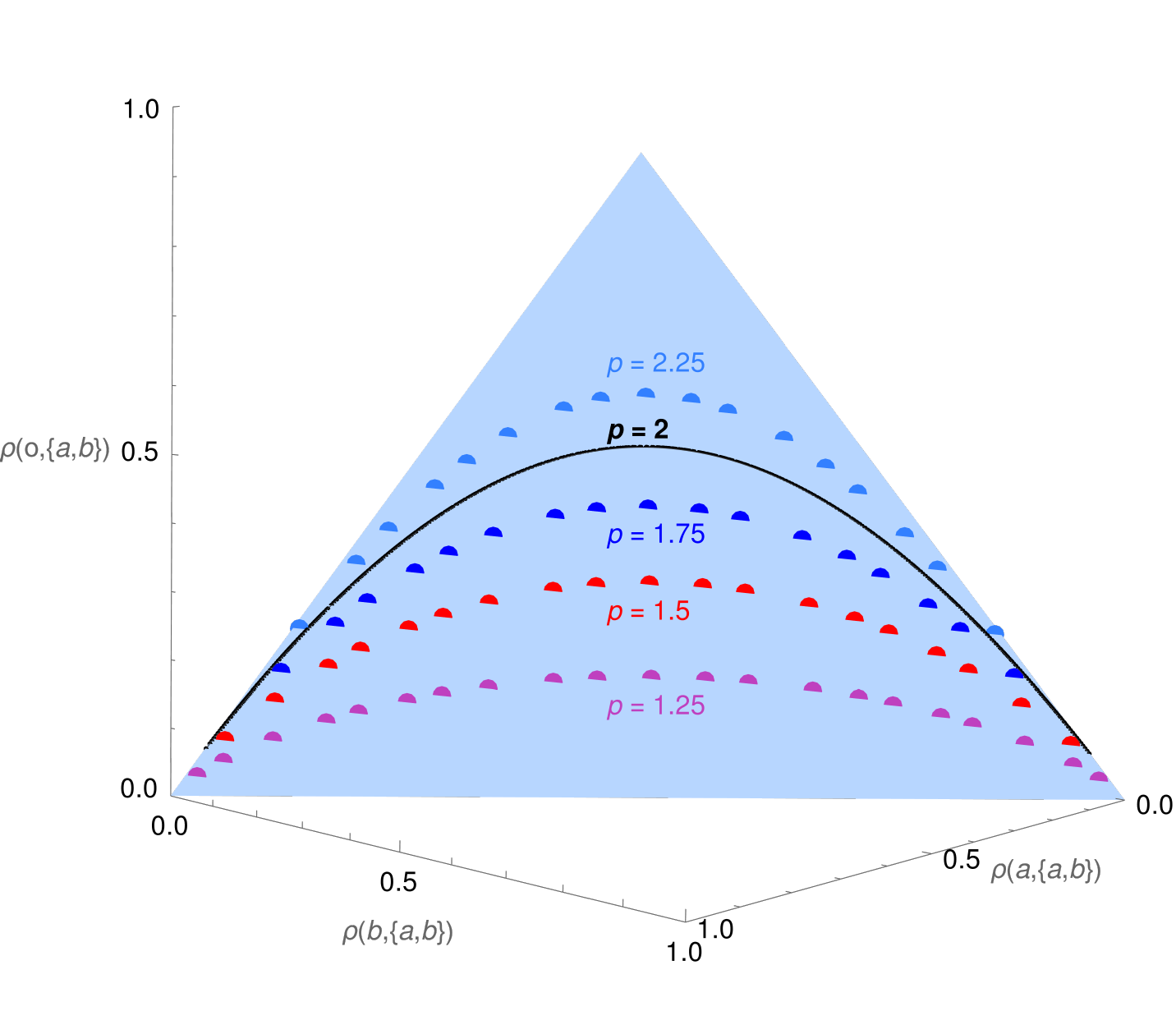}
	\label{fig:symmetry}
\end{figure}

\subsection{Three Empirically Supported Behavioural Predictions}

We proceed with an illustration of how a general $(u,D)$ 
model, or even the more structured power logit,  
make predictions that help explain intuitively the three  
empirically documented choice-deferral phenomena mentioned
in the Introduction. 
We start by noting that findings and arguments from the 
consumer-psychology literature reported in 
\citet{dhar97,sela-berger-liu09,scheibehenne-greifeneder-todd10}, 
among others, suggest that decision makers are sometimes more likely to 
avoid/delay choice when the feasible alternatives are perceived 
to be of similar value. 
The last authors noted, for example, that as the 
most attractive feasible options become more similar when new items 
are added to a menu, it can become more difficult for the decision 
maker to justify the choice of any particular option, 
which in turn would increase the likelihood of choice deferral. 
This is what we earlier referred to as \textit{``similarity-driven
deferral''}.
In the same direction, but focusing on response times rather than 
deferral decisions, \cite{bhatia-mullett18} recently reported 
evidence to suggest that choice between similarly attractive options 
is significantly correlated with longer response times.

Our next result shows how the power logit predicts such an effect.
More specifically, an interesting feature of this model is that its 
predicted probability of opting out at a menu as a function of the number 
of active-choice alternatives at that menu is bounded above in a simple way,
and that upper bound is attained precisely when \textit{all} feasible 
alternatives are of the same value.

\begin{prp}\label{bounds-quadratic}
	If $\rho$ is a power logit $(\widehat{u},p)$, then, for every menu $A$,
	
	\vspace{-5pt}
	
	$$
	\begin{array}{lllllll}
		\rho(o,A) 
		& \leq 	
		& 1 - |A|^{1-p}, 	
		& 						
		& 
		& \\
		\rho(o,A) 
		&  = 		
		& 1 - |A|^{1-p} 	
		& \Longleftrightarrow 	
		& \widehat{u}(a)=\widehat{u}(b) 
		& \text{for all } a,b\in A.
	\end{array}
	$$
\end{prp}
\begin{figure}[!htbp]
	\centering
	\caption{Maximum probability of deferring
		as a function of menu size in the power 
		logit model.}
	\includegraphics[width=0.65\textwidth]{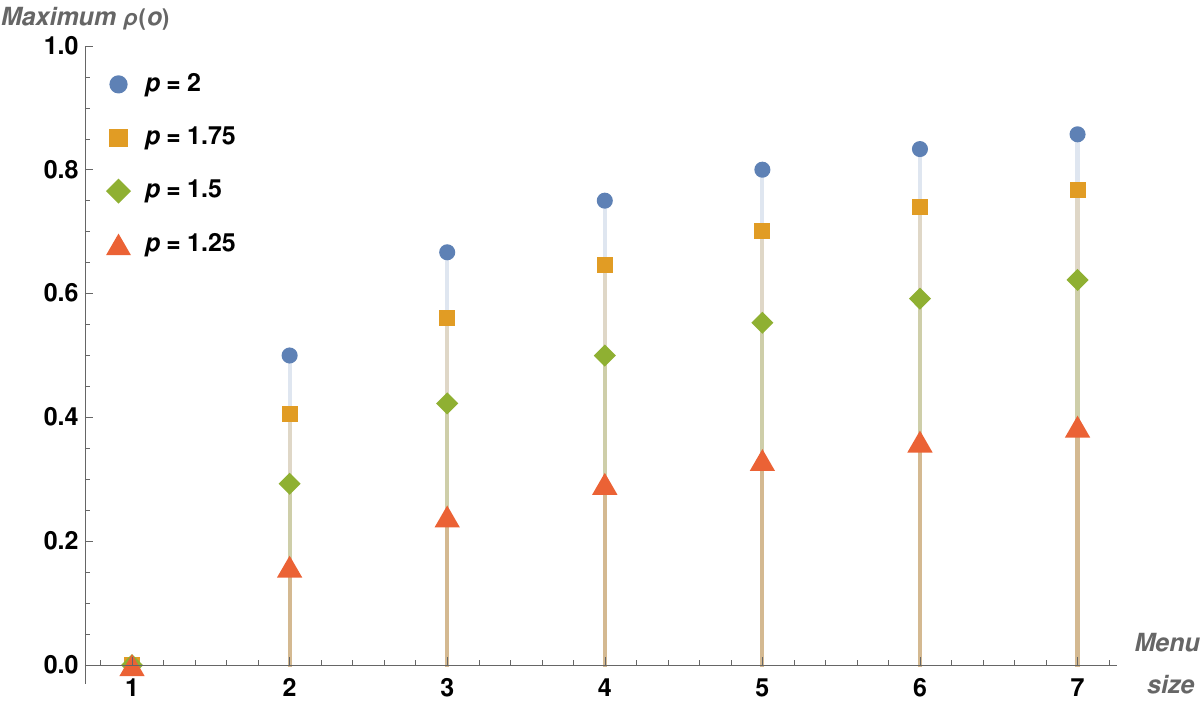}
	\label{deferralbounds}
\end{figure}
In this model, therefore, an agent's decision difficulty at a menu, 
as revealed by the deferral probability at that menu, 
is maximized when all feasible active-choice alternatives 
are equally desirable, and this maximum difficulty is increasing 
in proportion to the total number of such alternatives at a decreasing 
rate (Figure \ref{deferralbounds}).\footnote{A clarifying remark 
may be due at this point. Equal values (specifically, utilities) between
two or more alternatives are---in most of economic theory---associated 
with positive indifference, 
which in turn is interpreted as suggesting that the 
individual in question would be equally happy with any of 
these alternatives. 
By contrast, the influential \textit{drift diffusion model} 
in neuroeconomics \citep{krajbich-armel-rangel10,baldassi&etal,
fudenberg&newey&strack&strzalecki}, which originates 
in the psychology literature \citep{ratcliff&mckoon08}, 
and related experimental evidence that have been of 
increasing visibility and interest in the economics literature 
lately make the opposite predictions/observations. 
The latter in turn are broadly 
in line with the general predictions of
the power-logit model that we focus on in this paper. 
Considering the different motivations, methodological frameworks 
and intended interpretations in the two literatures, however, the
seeming discrepancy is in our view more an issue 
of semantics than it is one of substance. In any case, 
our use of the term ``value'' rather than ``utility'' 
in reference to the terms appearing in logit formulae 
is partly motivated by this issue.}

\begin{figure}[!htbp]
	\centering
	\caption{\centering Comparative statics in the quadratic logit 
		when one of the two alternatives becomes more attractive.}
	\includegraphics[width=1\textwidth]{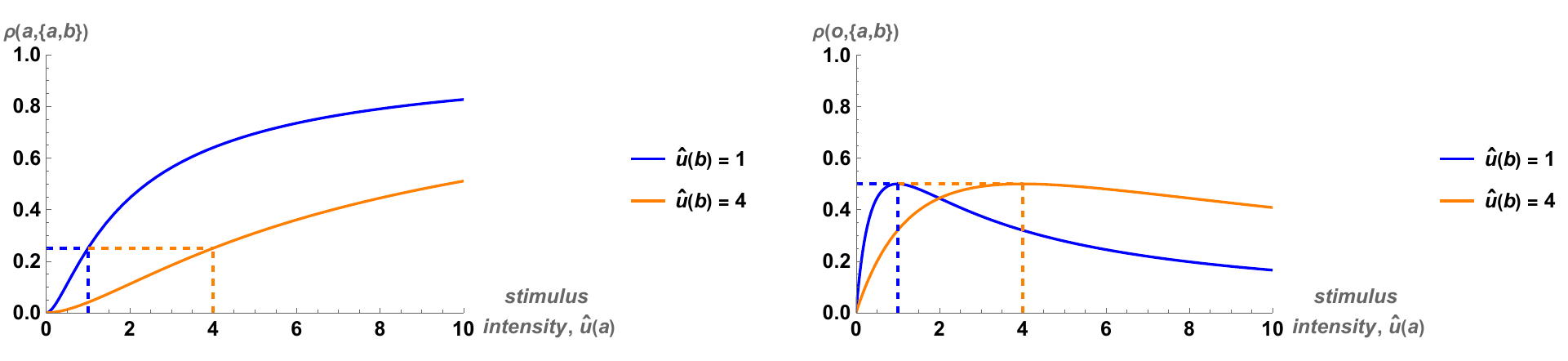}
	\label{fig:compstat-ql}
\end{figure}

We now turn to the power-logit model's comparative statics 
in the important class of binary menus. 
Figure \ref{fig:compstat-ql} illustrates, with a quadratic-logit example,  
the general pattern in the behavior of 
$\rho(a,\{a,b\})$ and $\rho(o,\{a,b\})$ as the stimulus intensity of 
$a$ changes while that of $b$ is held fixed. 
Interestingly, the monotonic increase of $\rho(a,\{a,b\})$ in 
$\widehat{u}(a)$ occurs at an increasing rate as this value approaches 
the $\frac{\widehat{u}(b)}{2}$ stimulus-intensity threshold from below 
than when $\widehat{u}(a)$ increases monotonically beyond 
$\frac{\widehat{u}(b)}{2}$. 
Intuitively, the inflection-point stimulus intensity value 
$\frac{\widehat{u}(b)}{2}$ that dissects $\rho(a,\{a,b\})$---viewed 
as a function of $\widehat{u}(a)$---into convex and concave 
regions suggests that marginal improvements in the appeal of $a$ lead 
to more rapid market share increases when this alternative is still 
``catching up'' with $b$ than when it has become sufficiently close 
to (or surpassed) it in attractiveness. 
On the other hand, $\rho(o,\{a,b\})$ is a strictly concave function of 
$\widehat{u}(a)$ and, consistent with Proposition \ref{bounds-quadratic}, 
attains its maximum value of $\frac{1}{2}$ when 
$\widehat{u}(a)=\widehat{u}(b)$. 
Thus, the model's novel prediction here, of relevance both from a 
consumer-welfare and a seller-profit perspective, 
is that minimal re-designing of a menu that consists of equally 
attractive alternatives is more likely to be effective at 
reducing opt-out behavior if one of the original alternatives 
\textit{becomes less rather than more appealing}, other things equal.

Now, since, as was discussed previously, it is not generally true that 
avoiding/deferring becomes more likely as menus expand even for 
monotonic decision-conflict logit models, 
it is naturally of interest to understand when, exactly, such behavior 
is to be expected in this environment. 
The general idea in answering this question 
is that, even if decision difficulty increases in absolute terms when new 
alternatives are introduced, when these new alternatives are sufficiently 
better than the pre-existing ones their added value will offset the 
elevated decision cost and will ultimately result in a higher probability 
of making an active choice at the larger menu. To state this more formally 
we will abuse notation slightly by letting
\begin{eqnarray}
	\label{notation-abuse}	u(S) & := & \sum_{s\in S}u(s)
\end{eqnarray}
stand for the total Luce value at menu $S\in\cm$.

\begin{obs}\label{roller-coaster}
	If $\rho=(u,D)$ is a decision-conflict logit, 
	then for any $A,B\in\cm$ such that $A\supset B$:
	\begin{eqnarray}
		\label{percentages} \rho(o,A) \leq \rho(o,B)
		& \Longleftrightarrow 
		& \underbrace{\frac{D(A)-D(B)}{D(B)}}_{\text{\shortstack{marginal cost\\ 
					from menu expansion}}} 
		\leq  
		\underbrace{\frac{u(A)-u(B)}{u(B)}}_{\text{\shortstack{marginal benefit\\ 
					from menu expansion}}}.
	\end{eqnarray} 
\end{obs}

This eloquent equivalence clarifies that the choice probability of opting out 
will decrease following menu expansion if and only if the marginal benefit 
of this expansion, as measured by the percentage increase in total value, 
exceeds its marginal cost, as measured by the percentage increase in 
decision complexity. This is a distinctive property of decision-conflict 
logit models. It clarifies that they do not belong to the random-utility 
class\footnote{See \textcite{apesteguia-ballester-lu},
\textcite{stoye19}, \textcite{strzalecki24} and references therein.} 
with an outside option, 
and enables them to explain simply the non-monotonic and dominance-driven 
effect that menu expansion has been known to exert on the probability of 
deferring \parencite{scheibehenne-greifeneder-todd10,chernev-bockenholt-goodman15},
which we earlier referred to as the 
\textit{``roller-coaster choice overload''} effect.  

\begin{table}[!htbp] 
	\centering
	\footnotesize
	\setlength{\tabcolsep}{5pt} 
	\renewcommand{\arraystretch}{1.2} 
	\caption{Illustration of ``roller-coaster'' 
	choice-overload predictions 
	with the quadratic logit.}
	\begin{tabular}{|c|c|c|c|c|}
		\hline
		Option 
		& $\widehat{u}$ 
		&  \small $\rho(\cdot,\{a,b\})$ 
		& \small $\rho(\cdot,\{a,b,c\})$ 
		& \small $\rho(\cdot,\{a,b,c,d\})$ \\
		\hline
		$a$
		& 10  
		& 0.980 
		& 0.250 
		& 0.007 \\
		\hline
		$b$
		& 0.1 
		& 0.001 
		& 0.000 
		& 0.000 \\
		\hline
		$c$
		& 9.9 
		& $-$  
		& 0.245 
		& 0.007 \\
		\hline
		$d$
		& 100 
		& $-$  
		& $-$  
		& 0.694 \\
		\hline
		$o$
		& $-$ 
		& \color{Mahogany}\textbf{0.019}  
		& $\nearrow$ \quad {\color{Mahogany}\textbf{0.505}} \quad $\searrow$ 
		& 
		\color{Mahogany}\textbf{0.292} \\
		\hline
	\end{tabular}
	\label{tab:roller-coaster}
\end{table}

Indeed, citing several studies in consumer psychology, 
the meta-analysis in \textcite{chernev-bockenholt-goodman15} notes that
\textit{``it has been shown that consumers are more likely to make a 
	purchase from an assortment when it contains a dominant option than 
	when such an option is absent''} (p. 338). This finding is important 
for the interpretation and policy responses to choice-overload phenomena 
of the kind that were first reported in \textcite{iyengar-lepper00}. 
To our knowledge, the decision-conflict logit is the first random-choice 
model that predicts this dominance-driven emergence and disappearance 
of choice-overload effects, and it does so without imposing 
any undesirability or inattention constraints. 
Table \ref{tab:roller-coaster} illustrates an example such effect 
that is predicted by the quadratic logit model.

\begin{figure}[!htbp]
	\centering 
	\caption{Illustration of relative-desirability effect predictions
		of the quadratic logit.}
	\vspace{1pt}
	\caption*{\footnotesize Case 1: the $\widehat{u}$-sum 
		is constant in all menus}
	\includegraphics[width=0.31\textwidth]{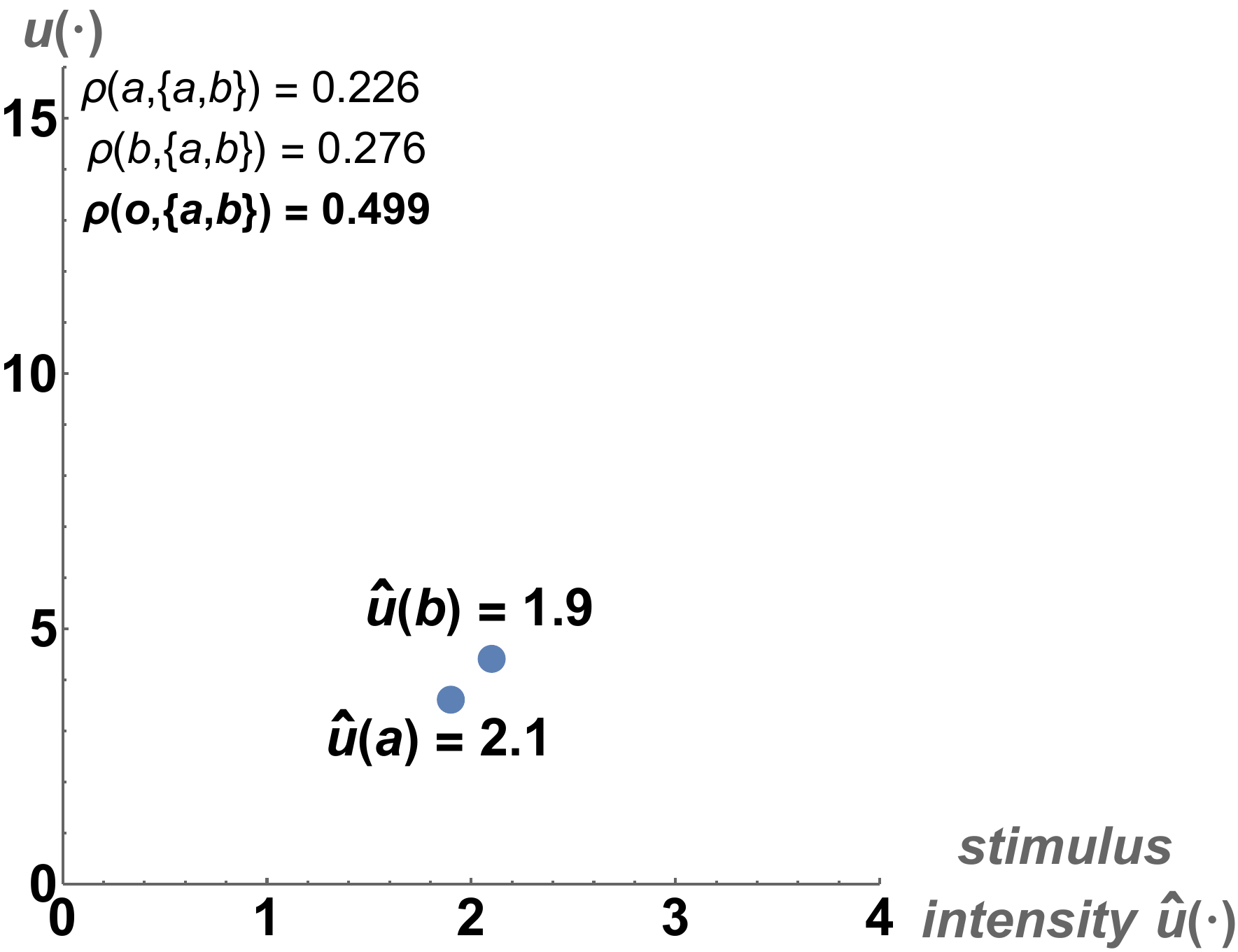}\;\;\;
	\includegraphics[width=0.31\textwidth]{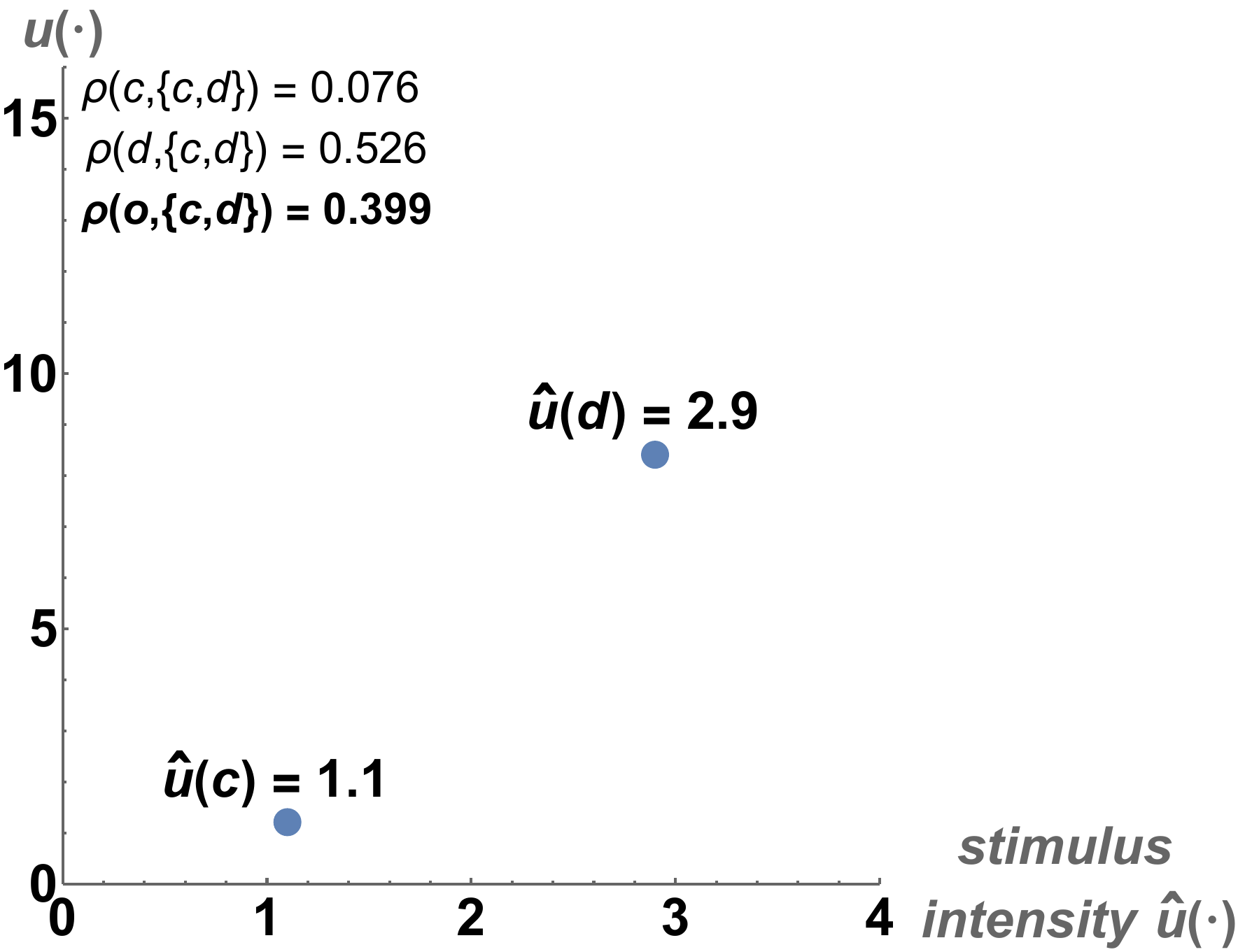}\;\;\;	
	\includegraphics[width=0.31\textwidth]{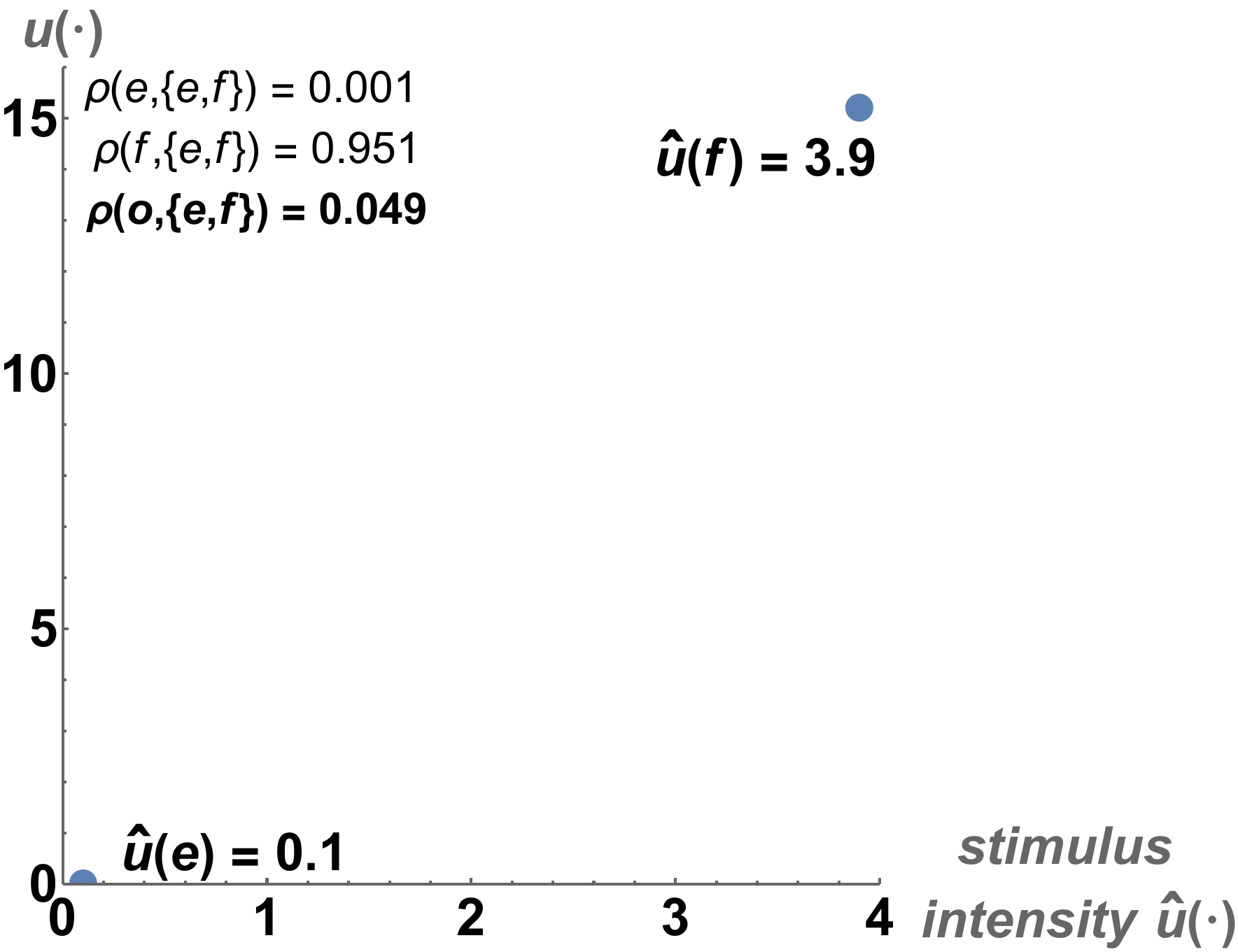}
	\caption*{\footnotesize Case 2: the $u$-sum is constant 
		in all menus}
	\vspace{10pt}
	\includegraphics[width=0.31\textwidth]{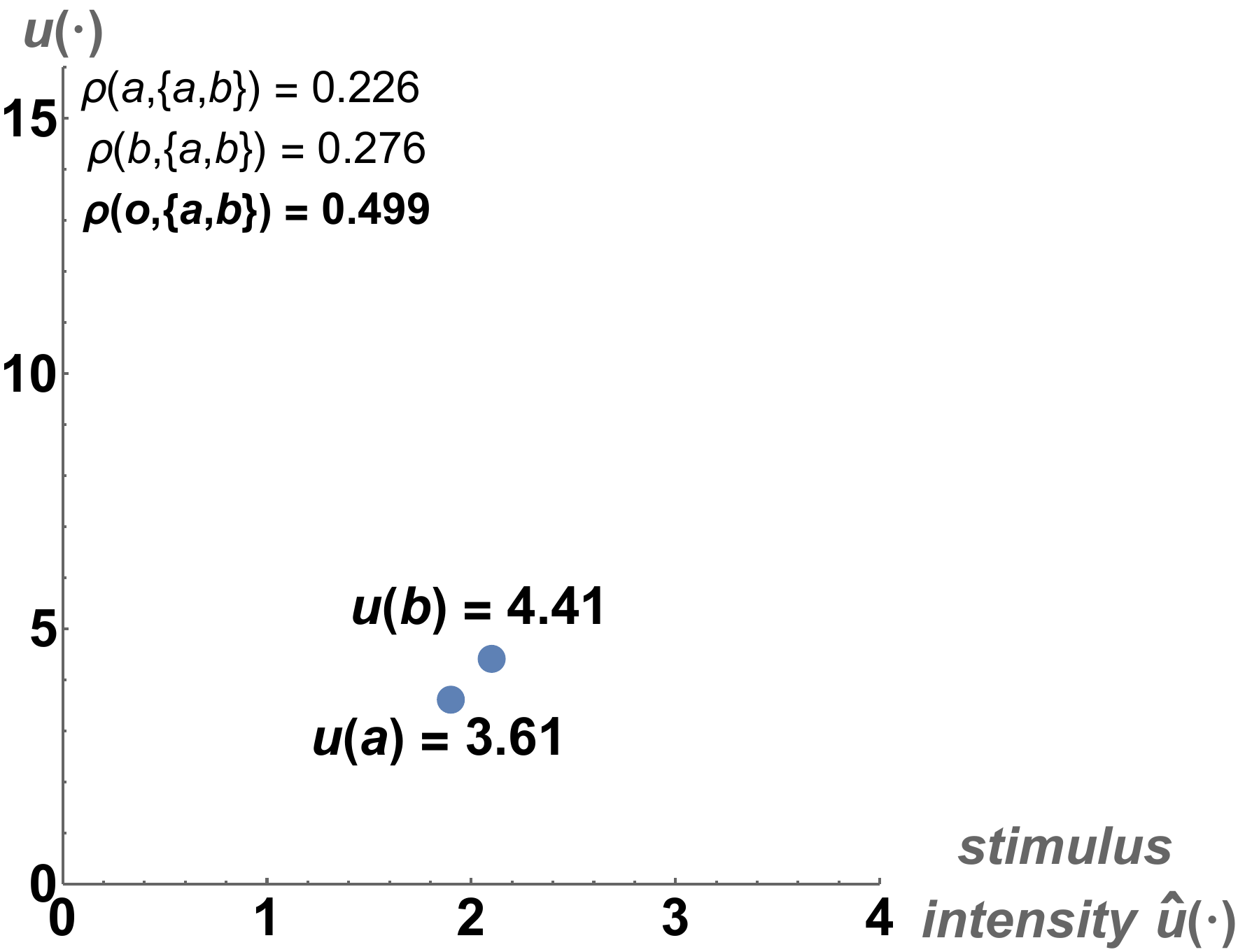}\;\;\;
	\includegraphics[width=0.31\textwidth]{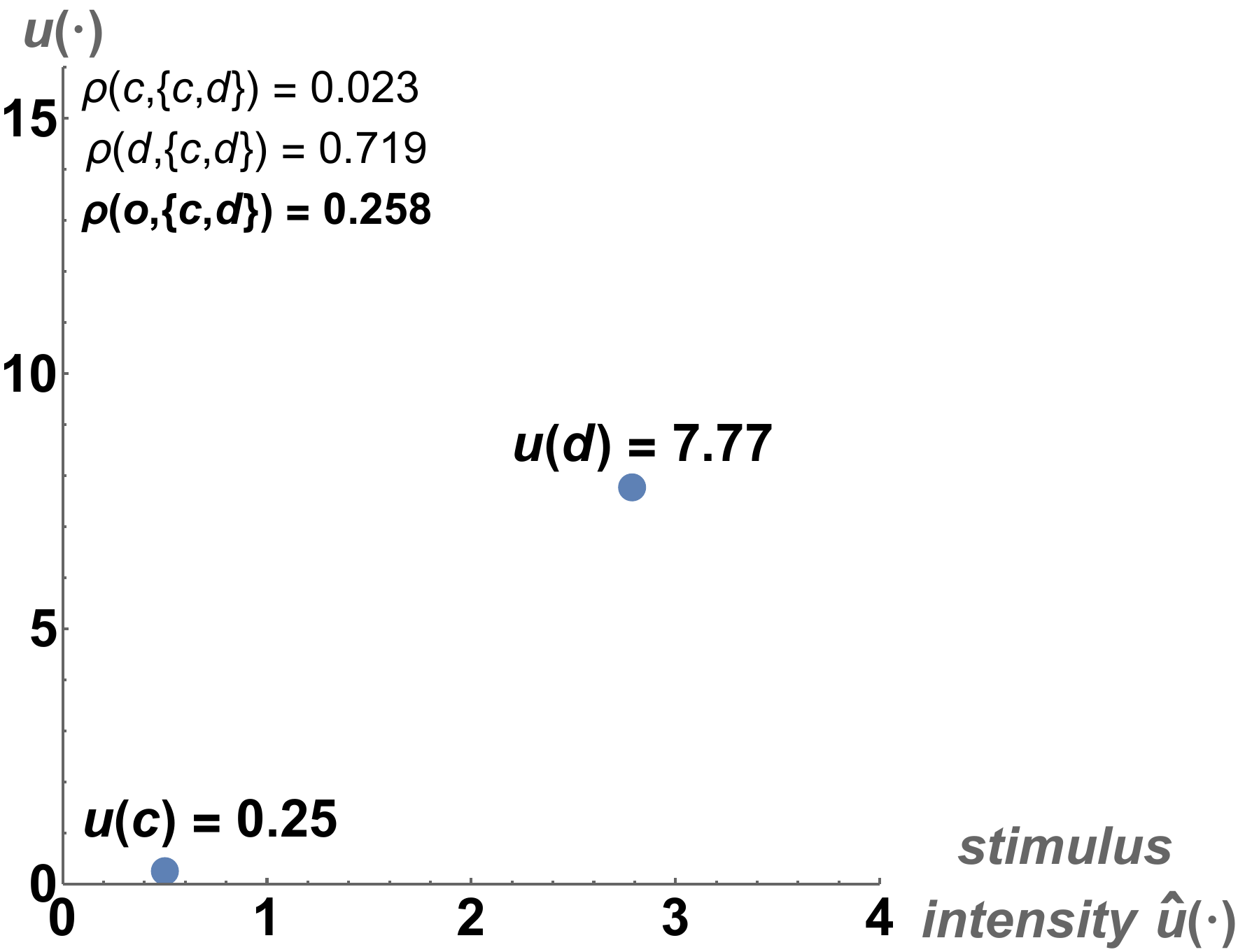}\;\;\;	
	\includegraphics[width=0.31\textwidth]{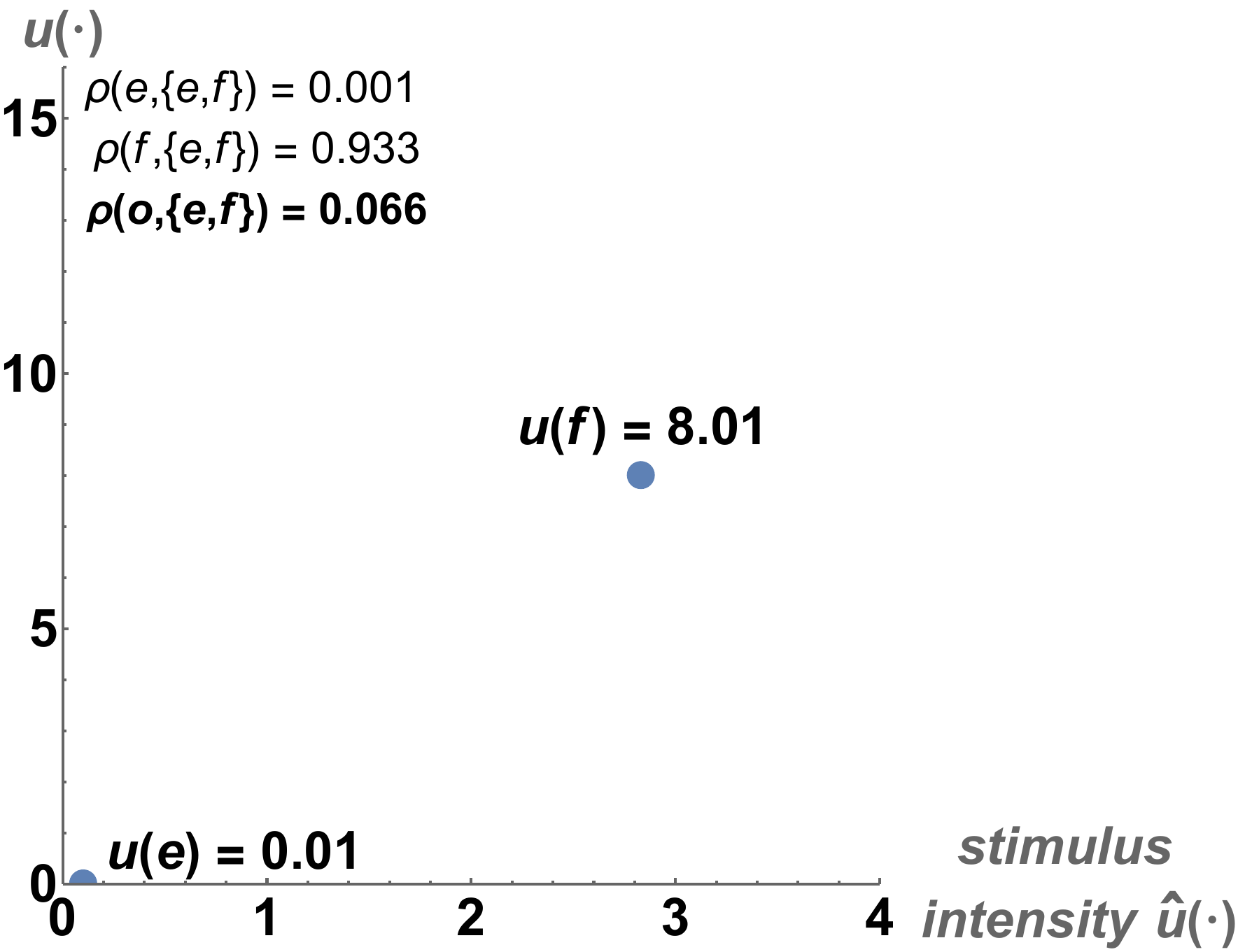}
	\label{fig:qlgraphs}
\end{figure}

Finally, as the next result establishes, the power-logit also
predicts another important choice-deferral phenomenon, known 
as the ``relative-desirability'' effect 
\parencite{dhar97,white-hoffrage-reisen15,bhatia-mullett16}. 
This refers to situations where, choosing the outside option
becomes more likely in binary menus as the available options
become more equally desirable, other things equal. 

\begin{prp}
	\label{prp:attract}
	If $\rho=(u,D)=(\widehat{u},p)$ is a power logit, then for any
	$a,b,c,d\in X$ where $\widehat{u}(a)+\widehat{u}(b) = \widehat{u}(c)+\widehat{u}(d)$
	or $u(a)+u(b) = u(c)+u(d)$ is true, the following is also true:
	\begin{eqnarray}
		\label{rel-attr1}  	\rho(o,\{a,b\}) > \rho(o,\{c,d\})
							& \Longleftrightarrow	&
							|\widehat{u}(a)-\widehat{u}(b)| < 
							|\widehat{u}(c)-\widehat{u}(d)|\\
		\label{rel-attr2}	& \Longleftrightarrow &
							|u(a) - u(b)| < |u(c) - u(d)|
	\end{eqnarray}
\end{prp}

\noindent 
This result, illustrated in Figure \ref{fig:qlgraphs}, 
is distinct from the similarity-driven deferral effect 
that was discussed in relation to Proposition 
\ref{bounds-quadratic} because it compares the probabilities 
of opting out at two distinct binary menus as a function 
of the absolute value/stimulus-intensity differences between 
the two active-choice alternatives, rather than focusing on 
when this probability is maximized within the same menu of any
size. It clarifies, indeed, that the model predicts 
relative-desirability effects irrespective of whether the 
stimulus-intensity absolute difference of the two alternatives
or that between their power-logit values---which emerge from the
stimulus-intensity values via the (convex) power 
transformation---is used to assess relative desirability. 

\section{Power-Logit Duopolistic Competition in Price and Quality}

We proceed with an illustration of the potential usefulness of the power-logit 
functional form in the analysis of oligopolistic markets when consumers 
potentially face comparison difficulties and may avoid/delay making an active 
choice.\footnote{\textcite{piccione-spiegler12}, \textcite{spiegler15}, 
	\textcite{bachi&spiegler} and \textcite{gerasimou&papi} 
	have recently suggested distinct approaches to study such markets.}
To this end, we consider a market where two profit-maximizing firms 
compete for a single consumer (equivalently, a unit mass of consumers) 
by offering a product that is differentiated in quality, 
$q_i$, and price, $p_i$. 
Producing a product of quality $q_i$ costs $q_i$ to firm $i=1,2$, 
while $0\leq q_i\leq p_i \leq I$ and $I>0$ denotes consumer income. 
Furthermore, a consumer's value from product $(q_i,p_i)$ coincides with 
that product's quality-price ratio:
\begin{eqnarray}
	\label{quality-price} u(q_i,p_i) & = & \frac{q_i}{p_i}.
\end{eqnarray}
This assumption further implies 
\begin{eqnarray}
	\label{utility-range} u(q_i,p_i) & \in & [0,1] 
\end{eqnarray}
for all $(q_i,p_i)$. 
Such a ``value-for-money'' specification imposes intuitive positive and negative 
dependences of $u$ on quality and price, respectively, 
with the former being linear and the latter strictly convex. 
Moreover, while identifying value with quality-price ratios 
as in \eqref{quality-price} rather than with quality-price differences 
$q_i-p_i$ appears to be a novel modelling assumption, 
it is consistent with some central implications of the 
behavioral choice model by \textcite{bordalloetal13} concerning consumer 
preferences for high quality-price ratio products, even though that model 
starts from very different primitives and features a quality-price 
difference value function instead.

The two firms choose their products' quality and price levels 
simultaneously and under complete information. 
The market share of product $(q_i,p_i)$ at menu/strategy profile 
$\big((q_1,p_1),(q_2,p_2)\bigr)$ is determined by the power logit model
\begin{eqnarray*}
	\rho\big((q_i,p_i),\{(q_j,p_j)\}_{j=1}^2\bigr) & = & 
	\left(\dfrac{\displaystyle\frac{q_i}{p_i}}{\displaystyle\frac{q_i}{p_i}
		+ \displaystyle\frac{q_j}{p_j}}\right)^s,
\end{eqnarray*}
where $s\geq 1$ and $s=1$ in the baseline special case where 
there is no decision difficulty.
Under the above assumptions, each firm $i=1,2$ solves
\begin{eqnarray}
	\label{profits} \max_{0\leq q_i\leq p_i\leq I} \pi_i(q_i,p_i) 
	& := & 
	(p_i-q_i)\cdot \rho\Big((q_i,p_i),\{(q_j,p_j)\}_{j=1}^2\Bigr)
\end{eqnarray}
The strategic trade-off in this model, which applies both 
when $s=1$ and $s>1$, is that each firm wishes to increase 
its quality/price ratio in order to expand its market share, 
while at the same time also wishing to decrease it in order 
to enlarge its profit markup.

Turning to consumer welfare, taking into account that decision conflict 
can potentially drive the consumer out of the market altogether, 
and that -by A3- this would be undesirable, we consider a 
utilitarian-like welfare measure that weighs the possible value levels 
at a given strategy profile by the probabilities that these values 
will actually be realized at that profile. We formalize this with 
the \textit{consumer welfare} function 
$W:\R^4_{++}\rightarrow [0,1]$ defined by
\begin{eqnarray*}
	W\big((q_i,p_i),(q_j,p_j)\bigr) & := & 
	\rho((q_i,p_i),\{(q_j,p_j)\}_{j=1}^2)\cdot u(q_i,p_i) +
	\rho((q_j,p_j),\{(q_j,p_j)\}_{j=1}^2)\cdot u(q_j,p_j).
\end{eqnarray*}
This welfare indicator may be particularly relevant in cases where consumer 
surplus is equilibrium-invariant, as will turn out to be the case 
in the present environment.\footnote{A related measure that identifies 
	welfare with the proportion of consumers who make an active choice 
	was studied in \textcite{spiegler15}, while \textcite{gerasimou&papi} 
	introduced an index that is similar to $W$ but features instead 
	the probability-weighted product variety that is associated 
	with a strategy profile.}

Perhaps surprisingly, this duopolistic model leads to the following simple
and intuitive equilibrium predictions:

\begin{prp}
	\label{competition}
	The power-logit equilibrium is 
	$(q_1^*,p_1^*)=(q_2^*,p_2^*)=\left(\dfrac{sI}{2+s},I\right)$
	and is associated with equilibrium expected profits
	$\pi_1^*=\pi_2^*=\dfrac{2^{1-s}}{2+s}I$ 
	and welfare $W^*=2^{1-2s}$.
\end{prp}		

Thus, although the equilibrium pricing strategy features full 
surplus extraction irrespective of the value of the hesitation/resampling 
parameter $s$, the equilibrium quality level increases in $s$ 
at the rate $\frac{s}{2+s}$. 
starting at the low of $\frac{I}{2}$ in the baseline case of 
logit market shares and no consumer hesitation ($s=1$), 
and approaching $I$ as $s$ becomes large.
An intuitive interpretation of this fact is that decision conflict 
inevitably introduces a third ``competitor'' into the market, 
the outside option, that becomes more ``powerful'' as $s$ grows.
The power logit predicts that the choice 
probability of the outside option goes down as the value of one 
of the two products is unilaterally increased, while the choice 
probability of the comparatively more appealing product 
simultaneously goes up during the process. This in turn creates 
incentives for each firm to unilaterally increase its quality 
level relative to the baseline logit case. But since increasing 
quality is costly, the above-mentioned strategic trade-off that 
is embedded in each firm's profit function eventually kicks 
in and halts this increase at the above symmetric-equilibrium level.

\begin{figure}[!htbp]
	\caption{\centering Power-logit equilibrium quantities in the
		duopolistic game as the power parameter varies\linebreak 
		\scriptsize{(income, $I$, is normalized to 1)}.}
	\includegraphics[width=1\textwidth]{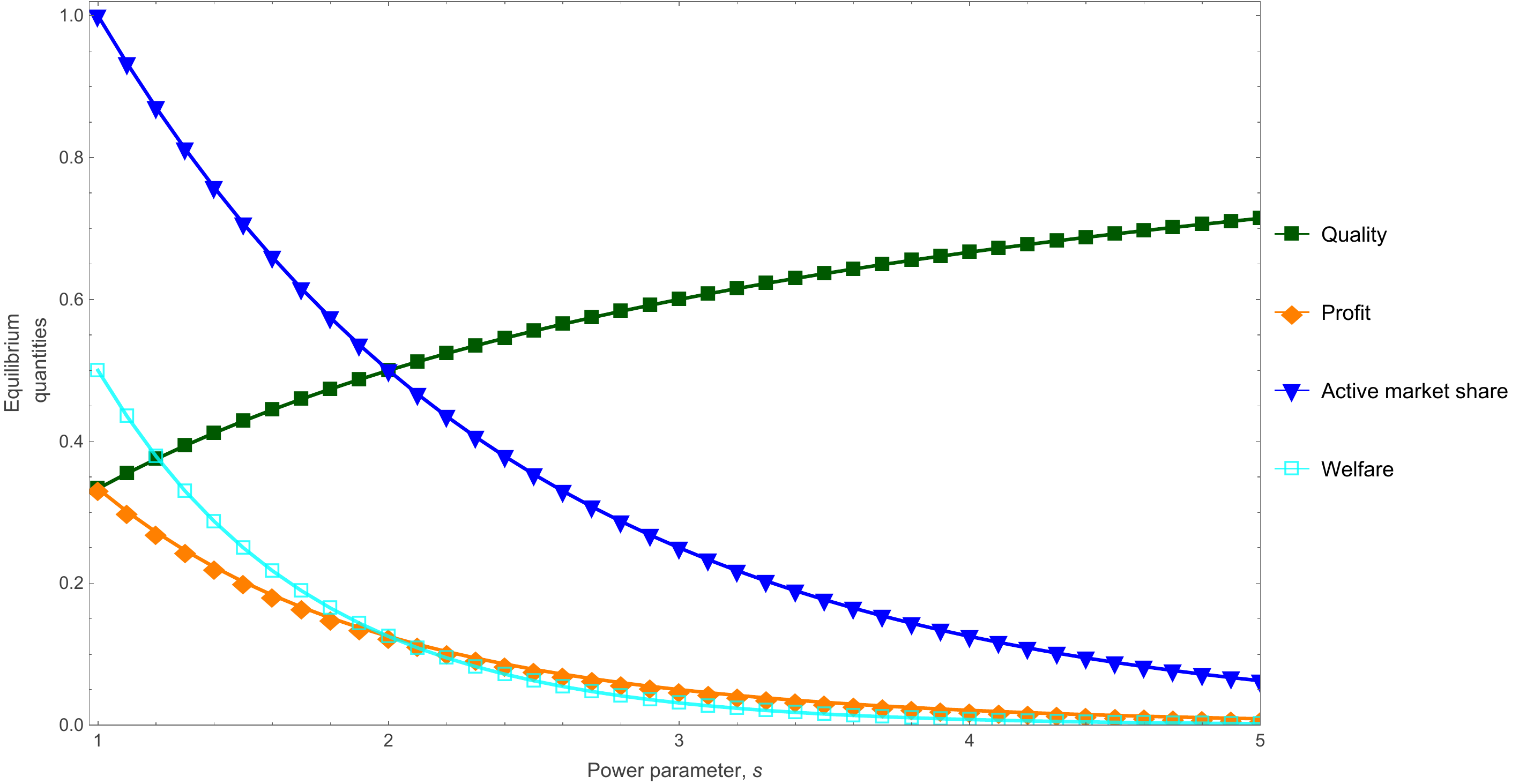}
	\label{fig:plogitEquilibrium}
\end{figure}

Notably, while consumer surplus is zero in equilibrium because each firm's 
profits turn out to be strictly increasing in its product's price, 
consumer welfare changes in an interesting way as $s$ varies. 
In particular, despite the increase in the attainable value level 
in equilibrium once firms best-respond to consumers' hesitation and resampling, 
welfare decreases in $s$. This decrease is caused by the fact that 
in the power logit with two equally attractive products the consumer 
is more/equally/less likely to defer than to make an active choice 
when $s>2/s=2/s<2$ and, conditional on doing the latter, 
equally likely to choose either of the two available products
(Proposition \ref{bounds-quadratic}). 
The implication of this in the present environment is that 
the higher value that the consumer receives 
\textit{in expectation} under the equilibrium with 
some decision conflict ($s>1$)
is not sufficiently high to offset the lower value that 
they receive \textit{with certainty} under the equilibrium 
with no conflict ($s=1$).
The firms' profits, finally, also decrease when consumers are 
hesitant relative to the case where there they are not. 
This large decrease is intuitive and contributed by the reduced probability 
of the consumer choosing either product, as well as by the reduction 
in the firms' profit margins that is brought about by the improvement in quality.
Figure \ref{fig:plogitEquilibrium} illustrates these facts graphically
when $I$ is normalized to 1.

\section{Econometric Estimation}

It is often the case in empirical applications 
that the choice frequencies available to the analyst are obtained from 
the choices made by a cross section of individuals who are 
presented with the same menu, rather than from a single decision maker's 
repeated choices at that menu. 
Random-utility based discrete choice estimation in those cases 
is often carried out under the assumption that the observable component 
of every individual's utility coincides, and that the error term 
in that model's formulation captures all individual heterogeneity 
that is unobserved to the analyst.
Adopting and adapting this assumption to our non-random-utility environment, 
in this section (and in Appendix A) we show how the other assumptions 
and formal argument that underpin the 
discrete-choice formulation of the logit model without 
an outside option that was pioneered by \textcite{mcfadden73} 
can be modified to arrive at a similar discrete-choice version of the 
quadratic- and power-logit models. 
It is worth remarking that, 
as we show in Section 5, the use of otherwise standard discrete-choice 
datasets is sufficient towards estimating these models, as long 
as they are obtained from a ``free choice'' decision environment, 
i.e. one where individuals could choose the no-choice outside option,
where the analyst observes both the active choices and those 
of the latter option.

\subsection{Discrete Choice with the Quadratic Logit Model}

We start by denoting the set of all quadratic-logit decision makers 
by $\{1,\ldots,n,\ldots,N\}$.
Keeping the menu $A:=\{a_1,\ldots,a_i,\ldots,a_k\}\subseteq X$ fixed throughout 
this and the next subsection, we proceed by recalling and breaking down 
the baseline assumptions of the discrete-choice formulation of the baseline 
logit in \eqref{logit-without} as follows:\\
1. \textit{Random utility} [structural assumption]:  
there is some function $u_n:X\rightarrow\R$ such that 
\begin{eqnarray}
	\label{mcfadden1} u_n(a_i) & = & v_n(a_i)+\epsilon_{ni}, 
\end{eqnarray}
where $v_n(a_i)$ and $\epsilon_{ni}$ are, respectively,
the deterministic (observable to the analyst) and stochastic (unobservable)
contributions to agent $n$'s utility from good $a_i$.
Denoting by $x_{ni}$ and $\beta$, respectively, two 
$m$-vectors of observable product/consumer characteristics 
and estimable coefficients that capture their relative importance via 
the relationship specified by some function  
$g:\mathbb{R}^m\times \mathbb{R}^m\rightarrow\mathbb{R}$, 
in applications it is assumed that 
\begin{eqnarray}
	\label{mcfadden1.1} 
	v_n(a_i)& \equiv & g(\beta; x_{ni})
\end{eqnarray}
and, often, that the dependence of $v_n$ on $\beta$ and $x_{ni}$ via $g$ is 
linear-additive:
\begin{eqnarray}
	\label{mcfadden1.5}
	g(\beta; x_{ni}) & = & \beta \cdot x_{ni},
\end{eqnarray}
where $\cdot$ denotes the inner product.\\
2. \textit{Random utility maximization} [behavioral assumption]:
for all $a_i\in A$, 
\begin{eqnarray}
	\label{mcfadden2} \rho_n(a_i,A) 
	& = & 
	Pr\big(u_n(a_i)\geq u_n(a_j) \text{ for all } j\leq k\bigr).
\end{eqnarray}
3. \textit{Gumbel noise} [distributional assumption]: 
the error term $\epsilon_{ni}$ is independently and identically 
distributed across $i$ according to the standard Gumbel density
\begin{eqnarray}
	\label{mcfadden3} f(\epsilon_{ni}) & = & 
	e^{-\epsilon_{ni}}e^{-e^{-\epsilon_{ni}}}. 
\end{eqnarray}
As has been widely known since the seminal 
contribution of \textcite{mcfadden73},\footnote{\textcite{luce&suppes} 
and, indeed, \textcite{mcfadden73} also credit 
Eric W. Holman and Anthony A. J. Marley with this discovery.} 
these assumptions jointly imply the analytically convenient and famous form
\begin{eqnarray}
	\label{discrete-choice} \rho_n(a_i,A) & = & 
	\frac{e^{v_n(a_i)}}{\sum\limits_{j=1}^ke^{v_n(a_j)}}.
\end{eqnarray}

We proceed by examining how the premises and conclusion 
of this classic discrete-choice logit model are affected 
and can be modified when we assume that decision maker $n$
uses the single but noisy value criterion captured by $u_n$ to sample 
the values of the alternatives in $A$ twice, 
as per the quadratic special case
of the power logit 
(focusing on the quadratic case here is done for simplicity 
of the exposition; we deal with the general case later).
To this end, and recalling the interpretation 
that was put forward at the beginning of Section 2.3,  
we first note that maintaining the 
additivity and linearity assumption implies that at the end of 
the second round of sampling the individual has perceived two values 
for each alternative $a_i\in A$,
\begin{eqnarray*}
	u^1_n(a_i) & = & v_n(a_i)+\epsilon_{ni}^1,\\
	u^2_n(a_i) & = & v_n(a_i)+\epsilon_{ni}^2.
\end{eqnarray*}

\noindent
These generally distinct values across the two rounds will vary according 
to the distribution of $\epsilon_{ni}$. Such multiplicity of value realizations 
in turn implies that each alternative $a_i\in A$ is ultimately associated 
with a vector of values $\big(u^1_n(a_i),u^2_n(a_i)\bigr)$. 
With utility now being vector-valued, however, the utility-maximization 
behavioral assumption that underpins \eqref{discrete-choice} is no 
longer applicable in an obvious way. 
To break this impasse we assume that the random utility maximization 
behavioral assumption is replaced by a \textit{dominance} assumption whereby
\begin{eqnarray}
	\label{dominance} \rho_n(a_i,A) & = & 
	Pr\big(u^l_n(a_i)\geq u^l_n(a_j) \text{ for all } j\leq k 
	\text{ and for } l=1,2\bigr).
\end{eqnarray}

\noindent
Turning, finally, to the modification of the distributional assumption 
\eqref{mcfadden3}, to make it operational in the quadratic-logit 
framework we assume that the random errors $\epsilon_{ni}^1$ and 
$\epsilon_{ni}^2$ are independent across all alternatives $i\leq k$ 
\textit{and} across the two sampling rounds $l\leq 2$.
As was also anticipated in the discussion of Section 3.1, 
this is indeed a demanding simplifying assumption that 
we hope future studies will be able to relax.

With these assumptions in place we can now write

\begin{eqnarray}
	\nonumber \rho_n(a_i,A)				
	& \equiv 
	& Pr(v_n(a_i)+\epsilon_{ni}^1\geq 
	v_n(a_j)+\epsilon_{nj}^1\; \forall\; j\neq i) \times 
	Pr(v_n(a_i)+\epsilon_{ni}^2\geq 
	v_n(a_j)+\epsilon_{nj}^2\; \forall\; j\neq i)\\
	\nonumber       					
	& = 
	& Pr(\epsilon_{nj}^1\leq v_n(a_i) +
	\epsilon_{ni}^1-v_n(a_j)\; \forall\; j\neq i) \times
	Pr(\epsilon_{nj}^2\leq v_n(a_i)+\epsilon_{ni}^2-
	v_n(a_j)\; \forall\; j\neq i)\\ 
	\nonumber       					
	& = 
	& \begin{split} 
		\int_{-\infty}^\infty\left(\prod_{j\neq i}e^{-e^{-(\epsilon_{ni}^1+
			v_n(a_i)-v_n(a_j))}}\right)
		e^{-\epsilon_{ni}^1}e^{-e^{-\epsilon_{ni}^1}}d\epsilon \\ 
		\times \hspace{100pt} \\
		\int_{-\infty}^\infty\left(\prod_{j\neq i}e^{-e^{-(\epsilon_{ni}^2+
			v_n(a_i)-v_n(a_j))}}\right)
		e^{-\epsilon_{ni}^2}e^{-e^{-\epsilon_{ni}^2}}d\epsilon\end{split}\\ 
	\label{quadratic-discrete}    					
	& = 
	& 
	\left(\frac{e^{v_n(a_i)}}{\sum\limits_{j=1}^{k} e^{v_n(a_j)}}\right)^2,
\end{eqnarray}
where each integral is $k$-dimensional, 
the first step makes use of the above behavioral, 
distributional and independence assumptions on $\epsilon_{ni}^l$, 
while the last step follows from the derivation of the 
discrete-choice logit 
[see, for example, \textcite[pp. 36-37 \& 74-75]{train09}].

An important difference between the discrete-choice 
version of the logit \textit{with} an outside option in \eqref{logit-with}
and its quadratic-logit counterpart is that in the former case 
the modeller specifies the value of that option 
\textit{exogenously} (see \cite{anderson_etal,hensher-rose-greene15}), 
whereas in the latter case this value emerges 
endogenously as a function of the observable characteristics of
all active-choice alternatives.
Indeed, assuming now---and in the remainder of this section and
the next---both \eqref{mcfadden1.1} and \eqref{mcfadden1.5}, 
upon rewriting \eqref{quadratic-discrete} as
\begin{eqnarray}
\label{dc-quad-logit}	
\rho_n(a_i,A) 
	& = 
	& \frac{e^{2\beta\cdot x_{ni}}}{\sum\limits_{j=1}^{k} e^{2\beta\cdot x_{nj}}+
		2 \sum\limits_{i\neq j}e^{\beta\cdot (x_{ni}+x_{nj})}}
\end{eqnarray}
one observes that 
\begin{eqnarray}
\label{quad-discr1}	u_n(a_i) & \equiv & e^{2\beta\cdot x_{ni}},\\ 
\label{quad-discr2}	D_n(A) & \equiv & 2 \sum_{i\neq j}e^{\beta\cdot (x_{ni}+x_{nj})}.
\end{eqnarray}
By contrast, in the baseline model we have
\begin{eqnarray}
	\label{base-discr1}	u_n(a_i) & \equiv & e^{\gamma\cdot x_{ni}},\\ 
	\label{base-discr2}	u_n(o)   & \equiv & e^{\gamma\cdot x_{no}},
\end{eqnarray}
where $x_{no}$ is set by the analyst.

In Appendix A we consider the general case of the 
power logit model, formulate its likelihood function,
and identify the first-order conditions on $p$
and $\beta$ for its maximization.

\section{Proof-of-Concept Empirical Illustration}

\subsection{Free Choices from 100 Menus with Two Films}

For our application we use the survey-experiment data with film choices that were 
collected by \textcite{bhatia-mullett16}. 
In that study, 58 subjects were initially asked to rate
from 1 (least desirable) to 9 (most desirable)\footnote{A typo in 
	\cite{bhatia-mullett16} erroneously suggests that the highest 
	rating was 7 instead.}  the 100 most voted-on (hence
most popular) films on the IMDB online platform (https://www.imdb.com)
at the time. Following that, subjects were presented with 100 distinct 
binary menus with films that were drawn from that list, with the respective
images presented side by side.  
In the free-choice treatment, subjects 
were asked to choose either the film positioned on the \textit{left} or 
on the \textit{right} of each menu, or to \textit{defer} the decision
(these choices were entered by clicking on the left, right and up keys,
respectively). 

Regarding the instructions that subjects received, the authors highlighted 
(p. 136) that these \textit{``were created to avoid any suggestion of an
explicit time limit (e.g. to suggest that participants should defer if 
they cannot decide quickly enough) or that deferral was a third comparable 
option (e.g. in the form of a status quo or default movie). More
specifically, the instructions stated that if participants preferred 
the movie on the left/right then they should press the left/right arrow. 
If they could not make a decision about which of the two movies they preferred 
then they should press the up arrow instead.''}
In the forced-choice treatment, the same 100 menus were presented but deferral
was not feasible. 

The study featured a within-subject design and subjects 
were randomly assigned to start the experiment in either of the two treatments.
There was no limit in the time subjects had available to make their $2\times100$
decisions.

\subsection{Analysis}

\noindent Although \textcite{bhatia-mullett16} focused mainly on the 
relationship between choice deferral and response times, 
they also reported on the relationship between 
ratings and active-choice probabilities \textit{conditional on an active choice	
being made}. Specifically, they found that the film with a higher rating, 
where relevant, is chosen 83\% of the time (p. 137). 
Enabled in this way by the theoretical analysis of the previous sections, 
our focus here instead 
is on the \textit{unconditional analysis} of the 
explanatory value of the subjects' own ratings 
on their subsequent active-choice \textit{and} deferral decisions, 
and on comparing the results from this analysis when it  
builds either on the baseline logit with an outside option or on 
the hereby proposed power and quadratic logit.\footnote{We recall that, 
as was clarified in Sections 2 and 3, the power logit 
and the baseline logit with an outside option are non-nested models.}
In particular, on each of the 100 binary menus in this dataset  
(for which, we recall, 58 observations are available) we estimate 
and compare the goodness of fit of the models that we lay out below.

\subsection{Multinomial Logit with a Fixed Outside Option}

In line with existing practices (see, for example, pp. 411-414 in 
\cite{hensher-rose-greene15}), 
to estimate this model we treat the outside option as an explicit alternative
with a fixed value that is common to all subjects.{\footnote{Under 
these two conditions the exact value of the outside option's ``rating'' is 
unimportant for this model's maximized log-likelihood and 
estimate of $\beta_1^A$, mattering only for the estimates of 
$\beta_0^{l,A}$ and $\beta_0^{r,A}$.}} 
Doing so leads to the following three-parameter multinomial logit specification:
\begin{eqnarray}
\label{dc-logit-1}	P_{n}^{ML}(l,A) & = & 
\dfrac{e^{\beta_0^{l,A}+\beta_1^{A}rat.Left_{n}}}{1 + 
	e^{\beta_0^{l,A}+\beta_1^{A}rat.Left_{n}}
	+ e^{\beta_0^{r,A}+\beta_1^{A}rat.Right_{n}}}\\
\label{dc-logit-2}	P_{n}^{ML}(r,A) & = & 
\dfrac{e^{\beta_0^{r,A}+\beta_1^{A}rat.Right_{n}}}{1 + 
	e^{\beta_0^{l,A}+\beta_1^{A}rat.Left_{n}}
	+ e^{\beta_0^{r,A}+\beta_1^{A}rat.Right_{n}}}\\
\label{dc-logit-3}	P_{n}^{ML}(o,A) & = & 
\dfrac{1}{1 + e^{\beta_0^{l,A}+\beta_1^{A}rat.Left_{n}}
	+ e^{\beta_0^{r,A}+\beta_1^{A}rat.Right_{n}}}
\end{eqnarray}
The left-hand-side terms denote the estimated probabilities of 
subject $n$ choosing ``left'', ``right'' or ``defer'' at 
binary menu $A$. 
On the right hand side, $\beta_1^A$ and $\beta_0^{l,A}$, $\beta_0^{r,A}$ 
are, respectively, the estimated slope and intercept coefficients 
at menu $A$. The former captures the effect that a unitary increase 
in subject $n$'s rating of the left (right) film--denoted here by 
$rat.Left$ ($rat.Right$)--has on the log-odds of choosing that 
film over deferring when the latter option's value is fixed.
The option-specific intercepts $\beta_0^{l,A}$ and $\beta_0^{r,A}$ 
on the other hand capture the log-odds of choosing, respectively,
the left and right film over deferring when the relevant film's 
rating is zero. Hence, including these terms in the estimation
is essential for otherwise the prediction would be equal choice 
probabilities for ``left'', ``right'' and ``defer'' 
if both films had a zero rating. This, in turn, would go against 
the model's treatment of the outside option as any other 
alternative that is more likely to be chosen as the other 
feasible options become worse.

\subsection{Multinomial Logit with a Randomly-Valued Outside Option}

We also consider the variant of the preceding model where, instead of assuming a fixed
common value (``rating'') for the outside option, we allow it to vary 
across subjects and menus by randomizing uniformly 
over the permissible rating values.\\

\subsection{Quadratic Logit}

As discussed in the previous subsection, estimating the quadratic logit
amounts to estimating the parameter $\gamma^A$ in
\begin{eqnarray*}
	P_n^{QL}(l,A) & = & 
	\left(\frac{e^{\gamma^{A}\cdot rat.Left}}{e^{\gamma^{A}\cdot rat.Left} 
	+ e^{\gamma^{A}\cdot rat.Right}}\right)^2\\
	P_n^{QL}(r,A) & = & 
	\left(\dfrac{e^{\gamma^{A}\cdot rat.Right}}{e^{\gamma^{A}\cdot rat.Left} + 
		e^{\gamma^{A}\cdot rat.Right}}\right)^2\\
	P_n^{QL}(o,A) & = & 1 - P_n^{QL}(l,A) - P_n^{QL}(r,A)
\end{eqnarray*}
There are some important differences between this model and the multinomial
logit with an outside option laid out above.
First, unlike that model, the quadratic logit does not include any 
intercept terms. 
This is in line with the theoretical predictions of the general version of 
this model (Proposition \ref{bounds-quadratic}), according to which all 
active-choice options are equally likely to be chosen when they have the same value. 
Including alternative-specific intercept terms here would go against this 
prediction as it would lead to generally distinct predicted probabilities
for the left and right film when their ratings are identically equal to zero.
Second, unlike $\beta^A$, the slope coefficient $\gamma^{A}$ here 
captures the log-odds of \textit{choosing one film over the other}
(in particular, \textit{not} of choosing one film over deferring) 
following a unitary change in the former film's rating. 
More specifically, given \eqref{luce-0}, \eqref{auxiliary1}, 
\eqref{auxiliary3} and \eqref{quad-discr1}, a more appropriate 
interpretation of this coefficient is that it captures the relevant 
change in the log-odds of choosing one film over the other 
following a unitary increase in the former's rating 
\textit{conditional on an active choice having been made}, while  
the unconditional change in these log-odds 
is obtained by multiplying them by $1-\rho(o,A)$.
By contrast, \eqref{dc-quad-logit} clarifies that 
the log-odds of choosing a film over deferring following 
a unitary increase in that option's rating is captured by 
$2\gamma^A$ instead.

\subsection{Power Logit}

Estimating this more general model now involves finding 
simultaneously optimal values for the slope coefficient 
$\theta^A$ \textit{and} the power parameter $p_A$ in 

\begin{eqnarray*}
	P_n^{PL}(l,A) & = & 
	\left(\frac{e^{\theta^{A}\cdot rat.Left}}{e^{\theta^{A}\cdot rat.Left}
	+ e^{\theta^{A}\cdot rat.Right}}\right)^{p_A}\\
	P_n^{PL}(r,A) & = & 
	\left(\dfrac{e^{\theta^{A}\cdot rat.Right}}{e^{\theta^{A}\cdot rat.Left} + 
	e^{\theta^{A}\cdot rat.Right}}\right)^{p_A}\\
	P_n^{PL}(o,A) & = & 1 - P_n^{PL}(l,A) - P_n^{PL}(r,A),
\end{eqnarray*}

\noindent
The parameter $\theta^A$ here admits an analogous interpretation to $\gamma^A$
in the quadratic logit, while the term $p_A\theta^{A}$ 
is interpretable as the effect that a unitary change in a film's rating 
has on the log-odds of choosing that film over deferring.

\subsection{Model Estimation and Goodness-of-Fit Summary Comparisons}

We perform a goodness-of-fit analysis and comparison of the four models
that aim to assess their explanatory and predictive performance separately 
on each of the 100 menus.\footnote{The 
	results presented in this subsection were obtained with code 
	written in the R programming language (\cite{baseR}, v4.5.2)
	with RStudio \parencite{Rstudio}, and utilising the 
	``mlogit'' \parencite{mlogit}, ``optimx'' \parencite{optimx}, 
	``plyr'' \parencite{wickham11} and ``tidyverse'' \parencite{tidyverse} 
	packages/libraries.}
To this end, we focus on the maximized log-likelihood value, the 
Akaike (AIC) and Bayesian (BIC) information criteria, 
and each model's proportion of correct predictions. 
In particular, denoting by $\widehat{L}_A$, $k$ and $N_A$, respectively, 
a model's maximized log-likelihood value at menu $A$, 
the number of its parameters and its sample size, recall that 
$AIC = 2k-2\log(\widehat{L}_A)$ and 
$BIC =  k\log(N_A)-2\log(\widehat{L}_A)$.
The value of $k$ is 3 for the two multinomial logit models 
with a fixed and random outside option, 2 for the power logit 
and 1 for the quadratic logit. 
The sample size is $N_A=58$ in all four models and for 
each one of the 100 menus.
In the prediction analysis we used the models' 100 menu-specific 
predicted choices per subject ($5800=100 \times 58$ in total) 
to subjects' actual choices at each menu.
A model was taken to make a correct prediction for a given subject 
at a given menu if it predicted a weakly highest choice 
probability for the option that was actually chosen by 
that subject in that menu. The same principle was applied
in a comprehensive cross-validation analysis where data 
from all but one of the 100 menus were pooled together 
and the resulting model estimates were used to make 
out-of-sample predictions at the excluded menu.

\begin{table}[!htbp]
	\centering
	\footnotesize
	\caption{\centering Goodness-of-fit comparison of the 
	four models' estimates at the 100 menus.} 
	\setlength{\tabcolsep}{5pt} 
	\renewcommand{\arraystretch}{1.4} 
	\makebox[\textwidth][c]{
\begin{tabular}{|l|c|c|c|c|cr|}
	\hline
	\multicolumn{1}{|l|}{\multirow{2}{*}{Model}} & 
	\multirow{2}{*}{Parameters} &
	\multirow{2}{*}{Log-likelihood} & 
	\multirow{2}{*}{AIC} & 
	\multirow{2}{*}{BIC} & 
	\multicolumn{2}{c|}{\multirow{2}{*}{Correct predictions}} 
	\\ &&&&&&
	\\
	\hline 
	\multirow{1.5}{*}{Logit with fixed and} & 
	\multirow{2}{*}{3}&
	\multirow{2}{*}{85} & 
	\multirow{2}{*}{75} & 
	\multirow{2}{*}{60} & 
	\multirow{2}{*}{\quad 2067} & 
	\multicolumn{1}{c|}{\multirow{2}{25pt}{35.6\%}} 
	\\
	\multirow{1}{*}{inferior outside option} 
	&&&&&&\\
	\hline
	\multirow{1.5}{*}{Logit with random} & 
	\multirow{2}{*}{3}&
	\multirow{2}{*}{8} & 
	\multirow{2}{*}{7} & 
	\multirow{2}{*}{6} & 
	\multirow{2}{*}{\quad 2112} & 
	\multicolumn{1}{c|}{\multirow{2}{25pt}{36.4\%}}\\
	\multirow{1}{*}{outside option} 
	&&&&&&\\
	\hline
	Power logit & 
	2& 
	7 & 
	16 & 
	22 & 
	\quad 2414 & 
	\multicolumn{1}{c|}{41.6\%} 
	\\
	\hline
	Quadratic logit & 
	1 & 
	0 & 
	2 & 
	12 & 
	\quad 1804 & 
	\multicolumn{1}{c|}{31.1\%\quad} 
	\\
	\hline
\end{tabular}
}
\vspace{-3pt}
\makebox[\textwidth][l]{\scriptsize Note: the random outside option in the 
second model was estimated on uniform-random integer values between 1 and 9.}
\label{tab:goodness-of-fit}
\end{table}

\begin{table}[!htbp]
	\centering
	\footnotesize
	\caption{\centering 
	Summary of cross-validation analysis.}
	\setlength{\tabcolsep}{5pt} 
	\renewcommand{\arraystretch}{1.4} 
	\begin{tabular}{|l|c|c|}
		\hline
	\multirow{3}{*}{Model}
	&\multirow{1.5}{*}{Menus with strictly}
	&\multirow{1.5}{*}{}\\
	&\multirow{1}{*}{more correct predictions}
	&\multirow{1}{*}{Average** parameter} \\
	&\multirow{-1.5}{*}{out-of-sample*}
	&\multirow{-1.5}{*}{estimates} \\
	\hline
	\multirow{1.5}{*}{Logit with fixed and} 
	& \multirow{2}{*}{68} 
	& $\overline{\beta}_0^L=-2.87$, $\overline{\beta}_0^R=-2.99$ \\
	\multirow{1}{*}{inferior outside option} 
	&
	&$\overline{\beta}_1=0.59$ \\
	\hline
	Power logit
	& 17
	& $\overline{p}=1.48$, 
	$\overline{\theta}=0.41$ \\
	\hline
	\end{tabular}
\vspace{-3pt}
\makebox[\textwidth][c]{\scriptsize *Both models predicted
correctly at 15 menus. **Across all leave-one-menu-out estimations.}
\label{tab:cross-validation}
\end{table}

Figure \ref{fig:corr_p_slope} plots the 100 pairs of power- and slope-parameter
estimates that emerge from the power-logit model. 
The mean, median and standard deviation of the $p$ estimates 
in those regressions are 1.51, 1.47 and 0.27, respectively. 
The slope-parameter estimates on the other hand
have a mean, median and standard deviation of 0.43, 0.40 and 0.15, suggesting 
that the effect of a one-unit increase in a film's rating is an approximately
53\% increase in the odds of choosing that film over the alternative.
For comparison, the mean/median and standard deviation in the slope estimates
corresponding to the baseline logit with a fixed outside option are 
0.58 and 0.15, respectively, pointing to an approximately 78\% increase
in the above-mentioned odds.

Interestingly, there is a negative correlation (Spearman $\rho=-0.33$) between
the $\widehat{p}$ and $\widehat{\theta}$ estimates in these data. 
The fact that $\widehat{p}$ tends 
to be lower at menus where $\widehat{\theta}$ is higher, 
however, can be interpreted 
intuitively through the lens of this model. 
Specifically, when $\widehat{p}$ is high, the deferral frequency 
also tends to be high. When deferrals are primarily caused by the relative 
undesirability of the two films, as per the logit with an 
outside option, a higher value of the slope parameter would be expected, 
in line with the 
$\overline{\beta}_1>\overline{\theta}$ finding. 
This is so because, in this model, 
the marginal effect of a unitary change 
in a film's rating is more likely to be high when both films 
have a low rating.  
But when deferrals are not primarily due to undesirability but, instead, 
are mainly caused by decision difficulty, then relatively low values of 
$\widehat{\theta}$ could be observed not because of low but 
because of \textit{similar} ratings, and by the harder comparison 
that such similarity entails.

\begin{figure}[!htbp]
	\centering
	\caption{\centering 
		Joint distribution of the power parameter in the 100 power-logit 
		regressions and the average absolute differences in ratings 
		at the respective menus.}
	\vspace{-40pt}
	\includegraphics[width=0.9\textwidth]{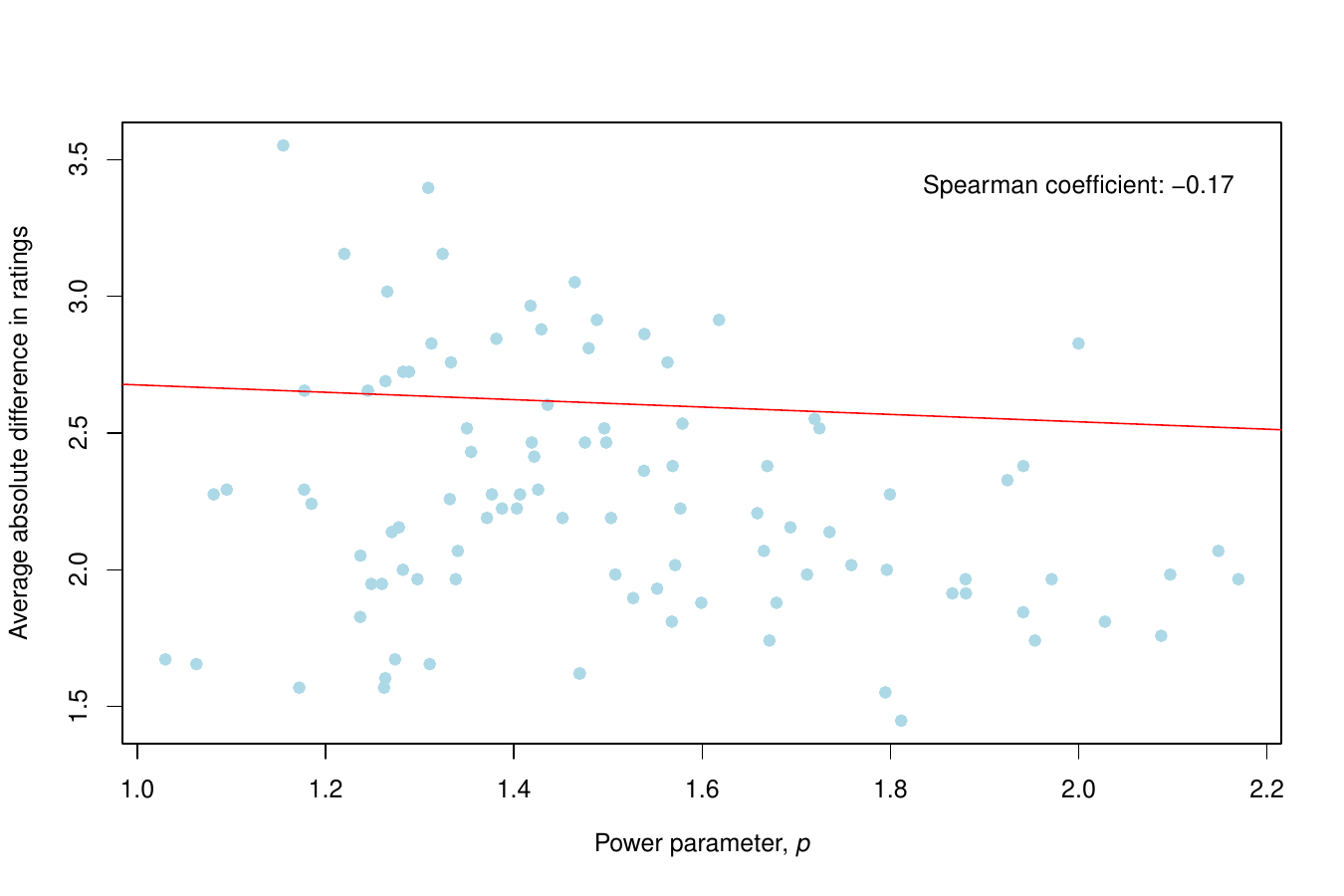}
	\label{fig:corr_p_average_rating_difference}
\end{figure}

The potential presence of such a channel is further supported 
by the negative correlation (Spearman $\rho=-0.17$) 
between the $\widehat{p}$ 
estimates and average (across subjects) absolute 
differences in ratings at the respective menus. 
The mean, median and standard deviation of this variable
at the 100 menus are 2.25, 2.21 and 0.45, respectively.
The bottom-right quarter of the scatter plot 
in Figure \ref{fig:corr_p_average_rating_difference} 
reveals the presence of 31 menus with an estimated $p$ 
in excess of its median value of 1.47 and an 
average absolute difference in ratings between the two 
films at each of these menus below its median of 2.25.
The mean and median estimates of the power-logit slope 
parameter $\widehat{\theta}$ at these 31 menus are 0.39, while 
the corresponding statistics in the remaining 69 menus
are 0.46 and 0.42. The difference in the distribution 
of $\widehat{\theta}$ between these two groups is statistically 
significant ($p=0.044$; two-sided Mann-Whitney test)
and corroborates this intuition and theoretical prediction.

We now turn to the results of the goodness-of-fit comparison that is 
summarized in Table \ref{tab:goodness-of-fit}.
The logit with an inferior outside option performs better than 
the other three models in most menus under each of the 
log-likelihood (85),  AIC (75) and BIC (60) criteria. 
Together, the power and quadratic logit provide the best fits 
under AIC and BIC in 18 and 34 menus, respectively, 
followed by the logit with a random outside option (7 and 6 menus). 
Under the proportion-of-correct-predictions criterion, 
on the other hand, the power logit performs better (41.6\%), 
followed by the baseline logit with a fixed or 
random outside option (35.6\% and 36.4\%, respectively) 
and by the quadratic logit (31.1\%).\footnote{The 5- and 4-percentage point 
differences in the rates of correct predictions between the power logit 
and the baseline logit with either an inferior or a 
random outside option are significant 
($p<0.001$ in both cases; $p$-values from 2-sided 
Fisher's exact tests).} 
Our cross-validation analysis (Table \ref{tab:cross-validation})  
focuses on the two models that the preceding analysis suggests are the 
descriptively leading ones. It reveals that 
the baseline logit---but not the power logit---makes 
correct out-of-sample predictions regarding the option that 
is most likely to be chosen in nearly two thirds of all cases.
Conversely, the power logit---but not the baseline logit---makes
correct out-of-sample predictions in nearly one fifth of all cases. 
Finally, both models make correct out-of-sample predictions 
15\% of the time.

\begin{figure}
	\centering
	\caption{Joint distribution of the power and slope parameters
		in the 100 power-logit regression estimates.}
	\vspace{-40pt}
	\includegraphics[width=0.9\textwidth]{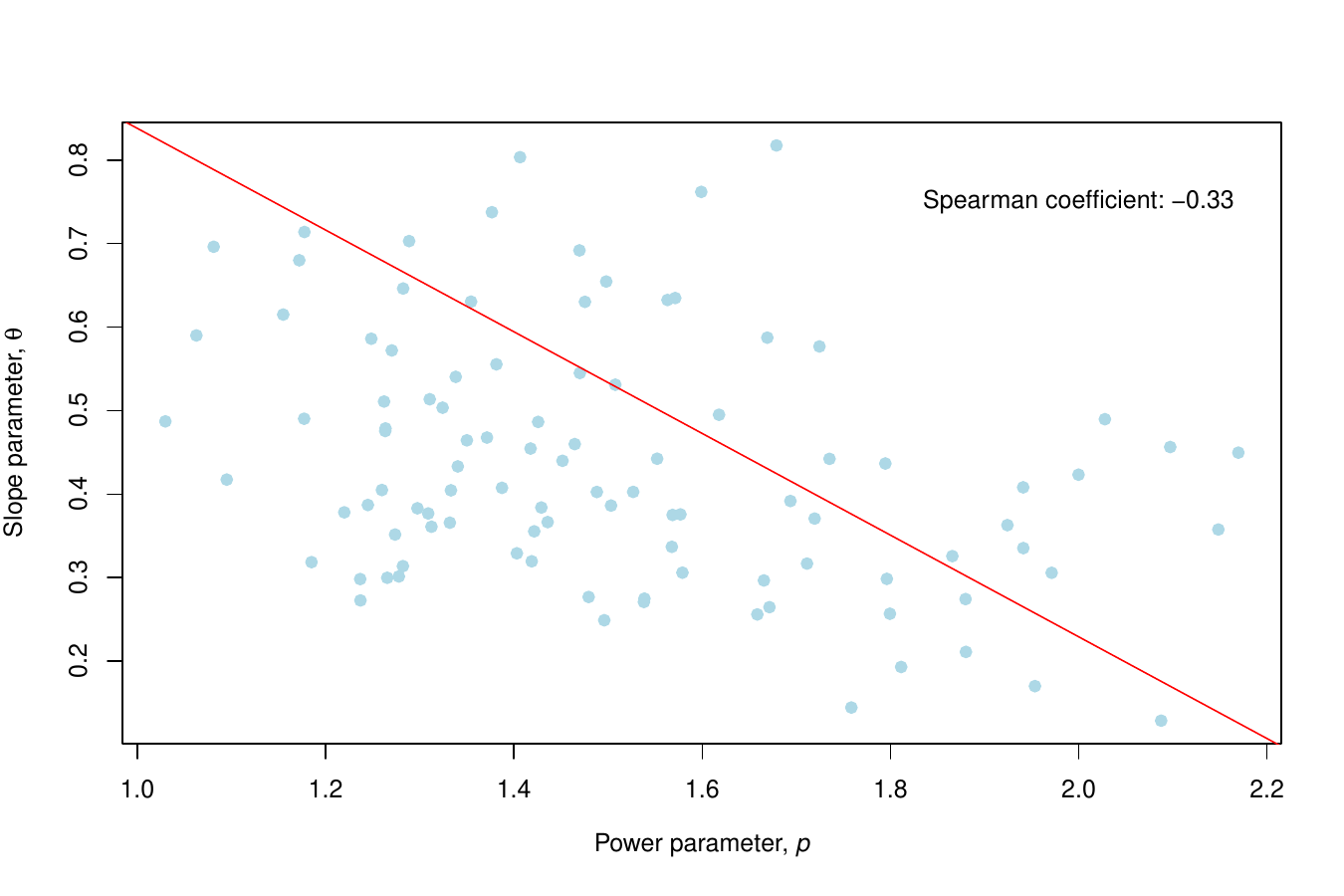}
	\label{fig:corr_p_slope}
\end{figure}

Further light on the relevance of the behavioral channel that 
was discussed earlier can now be shed by comparing the models' 
fit in those menus where the average film ratings are high and low. 
This is relevant because the mechanism underpinning the logit with a fixed
outside option suggests that choosing that option is more likely
when the average rating is low. 
Intuitively, therefore, we would expect this model to provide a better
fit in the latter group of menus than the power logit does.
To this end, we compare the two models' AIC and BIC scores 
in the two groups of 50 menus with above- and below-median average total
rating (the median value of this statistic is 11.44).
In line with this intuition, the baseline logit performs better 
than the power logit in a higher proportion of the 50 menus with 
a low average rating than those with a high such rating, 
under both criteria (AIC: 92\% vs 74\%; BIC: 84\% vs 60\%), 
with the difference in proportions being significant in both cases.\footnote{
The respective $p$-values from two-sided Fisher's exact tests are 
$p=0.03$ and $p=0.01$.}

These results suggest that the hereby proposed class of power-logit 
discrete-choice models with an endogenously determined 
menu-dependent value of the outside option can indeed 
provide meaningful explanatory gains relative to the baseline 
logit model with a fixed or random outside
option. Moreover, these explanatory gains often occur in those decision 
environments where intuition and the theoretical analysis of previous sections 
would suggest that the proposed model should indeed perform better. 
We hope that this illustration 
will be helpful to the experimenter or empirical researcher who is interested
in creating and analyzing similar free-choice datasets.

\section{Related Literature}

As was illustrated in the proof-of-concept empirical application
of the previous section, and as was also explained in Section 2, 
standard discrete choice models with an outside option that are based on 
random-utility maximization treat this option just 
like any other alternative and predict that it is more likely to be 
chosen when its utility is higher than that of all feasible 
\textit{active-choice} alternatives. 
\citet{anderson_etal} and \citet{hensher-rose-greene15}, for example, 
are textbook references that 
discuss this approach in detail.
The models that we study in this paper
differ radically from this (un-)desirability approach to modelling 
choice of the outside option, as they predict 
that every active-choice alternative is always chosen when 
it is the only feasible one (A3, Section 2). In addition, 
the more structured power-logit special case of this model predicts 
that the probability of opting out at larger menus increases 
as the feasible such alternatives become more equally appealing, 
in line with relevant empirical evidence.

Starting with \textcite{man&mar14}, several random choice models 
of limited attention---also logically distinct from the 
modelling framework that this paper proposes---have included 
an outside option as a model-closing assumption that requires 
this option to be chosen when no attention is paid to any of the 
feasible active-choice alternatives 
\citep{brady-rehbeck16,aguiar17,aguiar-boccardi-kashaev-kim23}.
Because of this assumption, deferring/opting out becomes less likely in these 
models as menus become bigger. \textcite{horan19} recently clarified  
how the deferral option can be removed from these models without affecting 
their general features and primary purpose, which is to explain 
active-choice decision making subject to cognitive/attention constraints. 

We proceed with a more detailed comparison of 
the logit with a general outside option 
that is formulated in \eqref{first} 
and the intuitive generalization of the classic \textit{nested logit} model 
\citep{ben-akiva73,mcfadden78} that was 
recently proposed and analysed in \citet{kovach-tserejigmid19}. 
The latter assumes that the set of alternatives $X$ can be partitioned 
into nests $X_1,\ldots,X_K$, and that there exist a non-negative
function $v^*$ on the collection $\bigcup_{i=1}^K2^{X_i}$
and a strictly positive function $u^*$ on $X$ such that, for all $A\in\cm$
and $a\in A\cap X_i$,
\begin{eqnarray}
	\rho(a,A) & = & 
	\dfrac{v^*(A\cap X_i)}{\displaystyle\sum_{j\leq K}v^*(A\cap X_j)}
	\dfrac{u^*(a)}{\displaystyle\sum_{b\in A\cap X_i}u^*(b)}
	\label{nsc}
\end{eqnarray}
That paper did not consider an outside option 
as it focused on explaining the kinds of canonical violations of A2 
that motivated the original development of nested logit 
as a generalization of baseline logit. 
Yet the version of \eqref{nsc} that is closest to \eqref{first} 
emerges when: (i) $X$ is expanded to $X\cup\{o\}$ and 
partitioned into the nests\footnote{See, for example, 
	the top branch of the auto-mobile choice model in Figure 1 of
	\citet{goldberg95}.} 
$\{X_1=\{o\}, X_2=X\}$; (ii)  
the collection of menus is 
$\cm^*:=\{A\cup\{o\}: \emptyset\neq A\subseteq X\}$.
In this case the choice probability of active-choice alternative $a$ 
at decision problem $A^*\in\cm^*$ reduces to
\begin{eqnarray}
	\rho(a,A^*) & = & 
	\dfrac{v^*(A^*\setminus\{o\})}{v^*(\{o\})+v^*(A^*\setminus\{o\})}
	\dfrac{u^*(a)}{\displaystyle\sum_{b\in A^*\setminus\{o\}}u^*(b)+u^*(o)}
	\label{nsc-o}				
\end{eqnarray}
Using the notational convention $A^*\equiv A\cup\{o\}$ for $A\in\cm$, 
a little algebra shows that \eqref{first} and \eqref{nsc-o} become 
equivalent if and only if $u(a)\equiv u^*(a)$ for all $a\in X$ 
and $D(A)\equiv \dfrac{v^*(\{o\})u^*(o)}{v(A)}$ for all $A\in\cm$.
Thus, unless $v^*(\{o\})=0$ or $u^*(o)=0$, 
equivalence between the logit with a general outside option and 
the generalized nested logit with a fixed  
outside option is possible \textit{only if} $D(A)>0$ for every $A\in\cm$.
The models that are representable as in \eqref{first} 
which do not impose this restriction at singleton menus.
Moreover, \eqref{nsc-o} has two additional degrees of freedom 
compared to \eqref{first}: one because $u^*$ takes 
$|X\cup \{o\}|=|X|+1$ values; 
and another because $v^*$ takes 
$|\mathcal{M^*}\cup\{o\}|=|\mathcal{M}|+1$ values.
Therefore, even when $D(A)>0$ for all $A\in\cm$, 
\eqref{nsc-o} is not uniquely recoverable from 
\eqref{first}. Thus, despite the structural similarity 
between \eqref{first} and \eqref{nsc-o}---which is perhaps best seen
by contrasting the two multiplicative terms in generalized nested 
logit with the corresponding ones in \eqref{luce-0}---the 
two models differ in some essential ways.

We note, finally, that also related to \eqref{first} but logically 
and interpretively distinct from it is the \textit{focal} 
logit model of \citet{kovach-tserenjigmid21} when a fixed 
outside option is introduced into the latter. 
That model's components comprise: 
(i) a menu-independent value function over alternatives
$u^{**}$ on $X\cup\{o\}$; 
(ii) a menu-dependent focus function $F$ that assigns 
a consideration set $F(A^*)$ to every problem $A^*\in\cm^*$; 
(iii) a menu-dependent focality bias function $\delta$ that gives 
a `value boost' to alternatives in $F(A^*)$.
Formally, the choice probability of active-choice
alternative $a\in A^*$ in this model is given by

\begin{eqnarray}
	\label{focal}
	\rho(a,A^*) & = & 
	\dfrac{u^{**}(a)\left(1+\delta(A)\times\mathds{1}\{a\in F(A^*)\}\right)}{
		\displaystyle\sum_{b\in A^*}u^{**}(b)\left(1+\delta(A)\times\mathds{1}\{b\in F(A^*)\}\right)},
\end{eqnarray}

\noindent
where $\mathds{1}\{\cdot\}$ is the indicator function.
Although \eqref{first} and \eqref{focal} are distinct, 
they intersect in the special case 
where $u(a)\equiv u^{**}(a)$ for all $a\in X$; 
$u^{**}(o)\equiv 1$; $F(A^*)\equiv \{o\}$ for all $A^*\in\cm^*$; 
and hence $\delta(A^*)\equiv D(A)-1$ for all $A^*\in\cm$ 
(recall that $A\equiv A^*\setminus\{o\}$).\footnote{The author is 
	grateful to Levent \"{U}lk\"{u} for alerting him to this connection 
	between \eqref{first} and \eqref{focal}.}
The last restriction implies $D(\{a\})>0$ for all $a\in X$, 
hence $\rho(a,\{a\})<1$. The special cases of \eqref{first}
that we studied here do not impose this restriction.

\section{Concluding Remarks}	

Understanding the ``easy'' and ``hard'' parts of people's preference comparisons,  
as these are revealed by their active-choice or choice-avoidance/delay decisions,  
is important \linebreak
methodologically and also for practical applications such as effective 
choice architecture. 
The present paper contributes in this respect by introducing the 
tractable power logit model and its quadratic-logit special cases. 
These models---both of which belong to a general class of logit models
with a context-dependent outside option---assume that people can avoid/delay 
making an active choice and are more likely to opt out in free-choice problems
when it is harder for them to identify a best alternative from those available to them. 
This prediction is supported empirically and differs from the predictions 
of existing models where the outside option is chosen due to the undesirability 
of all feasible alternatives, limited attention, or other sources of bounded-rational 
behavior. 
In conjunction with the insights from the relevant decision-making literature, 
our analysis suggests that decision-conflict logit models 
can help theoretical and applied empirical economists think 
formally and perhaps more realistically about non-strategic 
as well as strategic situations where decision makers: 
(i) are presented sufficiently small menus, so that limited-attention 
considerations are not pertinent; 
(ii) consider all feasible active-choice alternatives to be 
desirable/good enough, so that any one of them would be 
expected to be chosen if it were the only feasible item; 
(iii) find it difficult to compare these alternatives due to their complexity 
or due to potentially non-trivial trade-offs these generate; 
and (iv) are not forced to make an active choice.

\section*{Appendix: Proofs}

\noindent \textbf{\textit{Proof of Proposition \ref{characterization}.}} In the main text.\\

\noindent \textit{\textbf{Proof of Proposition \ref{bounds-quadratic}.}} 

For the second claim, suppose $\widehat{u}(a)=\widehat{u}(b):=c$ 
for all $a,b\in A$. 
By \eqref{power}, 
$\rho(o,A) = 1-
\sum\limits_{a\in A}\left(
\frac{\widehat{u}(a)}{\sum\limits_{b\in A}\widehat{u}(b)}
\right)^p
= 1-|A|\left(\frac{c}{|A|c}\right)^p =  
1-|A|^{1-p}$. Thus, $\rho(o,A)$ is independent of the specific 
$\widehat{u}$ values at $A$ whenever these values coincide. 
This readily implies that, viewed as the function 
\begin{eqnarray}
	\label{opt-out-function}
	\rho(o,A;\widehat{u}(a_1),\ldots,\widehat{u}(a_{|A|})) & = & 
	\dfrac{\left(\sum\limits_{i=1}^{|A|}\widehat{u}(a_i)\right)^p-
	\sum\limits_{i=1}^{|A|}\widehat{u}(a_i)^p}{\left(\sum\limits_{i=1}^{|A|}
	\widehat{u}(a_i)\right)^p},
\end{eqnarray} 
$\rho(o,A)$ has any $|A|$-vector of $\widehat{u}$ values $(c,\ldots,c)$ as a 
critical point that trivially satisfies both the first- and 
second-order conditions of local optimality. 
Yet, because the determinant of the Hessian matrix at any such point is zero, 
it is not immediately clear if this point is a local maximizer. 
To show that this is indeed so, by symmetry it suffices to 
consider marginal deviations in a single direction; 
say, an $\epsilon$ increase or decrease in $\widehat{u}(a_1)$.
Since $\widehat{u}(a_i)=\widehat{u}(a_j)\equiv c>0$, by assumption, 
this and \eqref{opt-out-function} yield
\begin{eqnarray*}
	\rho(o,A;c+\epsilon,c,\ldots,c) & = & 
	\dfrac{\left(|A|c+\epsilon\right)^p-
		(|A|-1)c^p-(c+\epsilon)^p}{\left(|A|c+\epsilon\right)^p}\\
	& = & 
	1 - \dfrac{(|A|-1)c^p+(c+\epsilon)^p}{\left(|A|c+\epsilon\right)^p},
\end{eqnarray*}
Suppose to the contrary that this weakly exceeds $1-|A|^{1-p}$. 
Without loss of generality, write $\epsilon:=mc$ for some small $m>0$
or $m<0$.
We have
\begin{eqnarray*}
	|A|^{1-p} & \geq & \dfrac{(|A|-1)c^p+(c+\epsilon)^p}{(|A|c+\epsilon)^p}\\
	& = & \dfrac{(|A|-1)c^p+(c+mc)^p}{(|A|c+mc)^p}
\end{eqnarray*}
To ease notation, write $n:=|A|$.
Rearranging, observe that the above is true if and only if 
$n^{1-p}(n+m)^pc^p \geq (n-1)c^p+(1+m)^pc^p$,
which in turn is true if and only if 
$n^{1-p}(n+m)^p \geq n-1 +(1+m)^p$.
Rearranging further, we get
$n\left(\dfrac{n+m}{n}\right)^p \geq n+1 + (m+1)^p$
from which we finally obtain 
$\left(1+\dfrac{m}{n}\right)^p - \dfrac{n+1+(m+1)^p}{n} \geq 0$
Taking the limit as $m\rightarrow 0$ and rearranging leads to
$n\geq n+2$, which is impossible. 
We have therefore established that the above critical point
is indeed a local maximizer of $\rho(o,A)$.

We proceed toward showing that it is in fact a global maximizer, 
thereby concluding the proof. 
To this end, notice first that $\rho(o,A)<1$, by strict positivity
of $\widehat{u}$.
Suppose to the contrary that there is a non-constant 
$|A|$-vector $(\widehat{u}(a_1),\ldots,\widehat{u}(a_{|A|})$ 
that satisfies the first-order conditions of optimality that 
are derived from \eqref{opt-out-function}.
Differentiating and rearranging pins down these conditions 
to
\begin{eqnarray}
	\widehat{u}(a_i^*) & = & 
	\left(\dfrac{\sum\limits_{j\neq i}\widehat{u}(a_j)}{\sum\limits_{j\neq i}
		\widehat{u}(a_j)^p}\right)^{\dfrac{1}{1-p}},
	\qquad i = 1,\ldots,|A|
\end{eqnarray}
Solving this system leads to
$\widehat{u}(a_1^*)=\widehat{u}(a_2^*)=\ldots=\widehat{u}(a^*_{|A|})$,
contradicting the supposed non-constancy of the postulated alternative local maximizer.
It follows that $\rho(o,A)$ is maximized at any constant $|A|$-vector only.
From this and the second claim that was established earlier it now 
follows that this maximum is indeed given by $1-|A|^{1-p}$, 
as per the first claim.
\sqbox\\

\noindent \textit{\textbf{Proof of Proposition \ref{prp:attract}}.}

To dispense with the absolute value sign, assume without loss of generality 
that 
$\widehat{u}(a)>\widehat{u}(b)$ and $\widehat{u}(c)>\widehat{u}(d)$.
We will first show that \eqref{rel-attr1} holds under either 
of the postulated conditions. 
Following that, we will show that \eqref{rel-attr1} 
$\Leftrightarrow$ \eqref{rel-attr2}, also under either 
condition.

Starting with \eqref{rel-attr1}, consider first the case where 
$\widehat{u}(a)+\widehat{u}(b)=\widehat{u}(c)+\widehat{u}(d)$.
Denote this common sum by $s$.
We have $\rho(o,\{a,b\})>\rho(o,\{c,d\}) \Leftrightarrow 
\frac{\widehat{u}(c)^p+\widehat{u}(d)^p}{s^p}
>\frac{\widehat{u}(a)^p+\widehat{u}(b)^p}{s^p}$. 
This is equivalent to 
\begin{eqnarray}
	\label{prp:attract-eq0}
	\widehat{u}(c)^p+\widehat{u}(d)^p & > & \widehat{u}(a)^p+\widehat{u}(b)^p
\end{eqnarray}
Suppose to the contrary that
\begin{eqnarray}
	\label{prp:attract-eq1}
	\widehat{u}(a) - \widehat{u}(b) & \geq &
	\widehat{u}(c) -\widehat{u}(d).
\end{eqnarray}
This and the postulated equality yield
$\widehat{u}(a)\geq \widehat{u}(c)$. 
Furthermore, this and \eqref{prp:attract-eq0} jointly imply
$\widehat{u}(a)>\widehat{u}(c)$ and $\widehat{u}(d)>\widehat{u}(b)$.
Thus,
\begin{eqnarray}
	\label{prp:attract-eq3}
	\widehat{u}(a)>\widehat{u}(c)>\widehat{u}(d)>\widehat{u}(b)
\end{eqnarray}
In view of \eqref{prp:attract-eq3}, observe that the terms 
$\frac{\widehat{u}(a)-\widehat{u}(c)}{\widehat{u}(a)-\widehat{u}(b)}$ and
$\frac{\widehat{u}(c)-\widehat{u}(b)}{\widehat{u}(a)-\widehat{u}(b)}$ 
are convex weights. Hence,
since $\widehat{u}(\cdot)\mapsto\widehat{u}(\cdot)^p$ is a strictly convex function,
we have
\begin{eqnarray}
	\nonumber
	\left(\frac{\widehat{u}(a)-\widehat{u}(c)}{\widehat{u}(a)-\widehat{u}(b)}\right)\widehat{u}(b)^p
	+
	\left(\frac{\widehat{u}(c)-\widehat{u}(b)}{\widehat{u}(a)-\widehat{u}(b)}\right)\widehat{u}(a)^p
	& > & \left[\left(\frac{\widehat{u}(a)-\widehat{u}(c)}{\widehat{u}(a)-\widehat{u}(b)}\right)\widehat{u}(b)
	+ \left(\frac{\widehat{u}(c)-\widehat{u}(b)}{\widehat{u}(a)-\widehat{u}(b)}\right)\widehat{u}(a)\right]^p\\
	\label{prp:attract-eq6} & = & \widehat{u}(c)^p,\\
	\nonumber
	\left(\frac{\widehat{u}(a)-\widehat{u}(d)}{\widehat{u}(a)-\widehat{u}(b)}\right)\widehat{u}(b)^p
	+
	\left(\frac{\widehat{u}(d)-\widehat{u}(b)}{\widehat{u}(a)-\widehat{u}(b)}\right)\widehat{u}(a)^p
	& > & \left[\left(\frac{\widehat{u}(a)-\widehat{u}(d)}{\widehat{u}(a)-\widehat{u}(b)}\right)\widehat{u}(b)
	+ \left(\frac{\widehat{u}(d)-\widehat{u}(b)}{\widehat{u}(a)-\widehat{u}(b)}\right)\widehat{u}(a)\right]^p\\
	\label{prp:attract-eq7} & = & \widehat{u}(d)^p
\end{eqnarray}
Adding \eqref{prp:attract-eq6} to \eqref{prp:attract-eq7} and recalling that
$\widehat{u}(a)+\widehat{u}(b)=\widehat{u}(c)+\widehat{u}(d)=s$ yields 
\begin{eqnarray*}
	\widehat{u}(c)^p + \widehat{u}(d)^p	& < & 
	\left(\frac{2\widehat{u}(a)-\widehat{u}(c)-\widehat{u}(d)}{\widehat{u}(a)-\widehat{u}(b)}\right)\widehat{u}(b)^p
	+
	\left(\frac{\widehat{u}(c)+\widehat{u}(d)-2\widehat{u}(b)}{\widehat{u}(a)-\widehat{u}(b)}\right)\widehat{u}(a)^p
	\\
	& = & 
	\left(\frac{2\widehat{u}(a)-s}{\widehat{u}(a)-\widehat{u}(b)}\right)\widehat{u}(b)^p
	+
	\left(\frac{s-2\widehat{u}(b)}{\widehat{u}(a)-\widehat{u}(b)}\right)\widehat{u}(a)^p\\
	& = & 
	\widehat{u}(b)^p + \widehat{u}(a)^p,
\end{eqnarray*}
which contradicts \eqref{prp:attract-eq0}. 
Thus, 
\begin{eqnarray}
	\label{prp:attract-eq4}
	\widehat{u}(a)-\widehat{u}(b) & < &
	\widehat{u}(c)-\widehat{u}(d)
\end{eqnarray}
holds. Conversely, suppose \eqref{prp:attract-eq4} is true. 
This and the postulated equality together imply 
\begin{eqnarray}
	\label{prp:attract-eq20}
	\widehat{u}(c)>\widehat{u}(a)>\widehat{u}(b)>\widehat{u}(d)
\end{eqnarray}
Applying the preceding convexity argument using \eqref{prp:attract-eq20} 
yields \eqref{prp:attract-eq0}, thereby completing the proof 
that \eqref{rel-attr1} holds under the first postulate.

We now show that \eqref{rel-attr1} is true when $u(a)+u(b)=u(c)+u(d)$ or, equivalently,
\begin{eqnarray}
	\label{prp:attract-eq8}
	\widehat{u}(a)^p+\widehat{u}(b)^p
	& =
	& \widehat{u}(c)^p+\widehat{u}(d)^p
\end{eqnarray}
holds instead. Let $t$ denote this common sum.
We have $\rho(o,\{a,b\})>\rho(o,\{c,d\}) \Leftrightarrow 
\frac{\widehat{u}(c)^p+\widehat{u}(d)^p}{\big(\widehat{u}(c)+\widehat{u}(d)\bigr)^p}
>\frac{\widehat{u}(a)^p+\widehat{u}(b)^p}{\big(\widehat{u}(a)+\widehat{u}(b)\bigr)^p}
\Leftrightarrow 
\frac{t}{\big(\widehat{u}(c)+\widehat{u}(d)\bigr)^p}
>\frac{t}{\big(\widehat{u}(a)+\widehat{u}(b)\bigr)^p}
\Leftrightarrow
\big(\widehat{u}(a)+\widehat{u}(b)\bigr)^p
>
\big(\widehat{u}(c)+\widehat{u}(d)\bigr)^p
$.
This is true if and only if 
$\widehat{u}(a)+\widehat{u}(b) >
\widehat{u}(c) + \widehat{u}(d)$,
which is equivalent to
\begin{eqnarray}
	\label{prp:attract-eq9} 
	\widehat{u}(a)-\widehat{u}(c) 
	& > 
	& \widehat{u}(d) - \widehat{u}(b)
\end{eqnarray}
Suppose to the contrary that 
\begin{eqnarray}
	\label{prp:attract-eq10} 
	\widehat{u}(a)-\widehat{u}(b)
	& \geq 
	& \widehat{u}(c)-\widehat{u}(d)
\end{eqnarray}
From \eqref{prp:attract-eq9} and \eqref{prp:attract-eq10} we get
$\widehat{u}(a)>\widehat{u}(c)$ and $\widehat{u}(b)<\widehat{u}(d)$. 
Thus, 
\begin{eqnarray}
	\label{prp:attract-eq11}
	\widehat{u}(a)>\widehat{u}(c)>\widehat{u}(d)>\widehat{u}(b)>0
\end{eqnarray}
By \eqref{prp:attract-eq9}, \eqref{prp:attract-eq10} and convexity 
of $\widehat{u}(\cdot)\mapsto\widehat{u}(\cdot)^p$ we have
\begin{eqnarray*}
	\widehat{u}(a)^p-\widehat{u}(c)^p
	& > &
	\widehat{u}(d)^p-\widehat{u}(b)^p,
\end{eqnarray*}
which contradicts \eqref{prp:attract-eq8}. 
Hence, \eqref{prp:attract-eq4} holds.
Conversely, suppose \eqref{prp:attract-eq4} is true
and assume to the contrary that \eqref{prp:attract-eq9} is violated, i.e.
\begin{eqnarray}
	\label{prp:attract-eq12}
	\widehat{u}(a)-\widehat{u}(c) 
	& \leq 
	& \widehat{u}(d) - \widehat{u}(b)
\end{eqnarray}
Rearranging \eqref{prp:attract-eq9},
\begin{eqnarray}
	\label{prp:attract-eq13}
	\widehat{u}(a)-\widehat{u}(c) & < &
	\widehat{u}(b)-\widehat{u}(d)
\end{eqnarray}
By \eqref{prp:attract-eq12} + \eqref{prp:attract-eq13}
we obtain $\widehat{u}(a)<\widehat{u}(c)$. 
This and \eqref{prp:attract-eq8} in turn imply $\widehat{u}(b)<\widehat{u}(d)$.
Hence,
\begin{eqnarray}
	\label{prp:attract-eq14} 
	\widehat{u}(c)>\widehat{u}(a)>\widehat{u}(b)>\widehat{u}(d)>0
\end{eqnarray}
By	\eqref{prp:attract-eq12} we have
\begin{eqnarray}
	\label{prp:attract-eq15}
	\widehat{u}(c)-\widehat{u}(a) 
	& \geq 
	& \widehat{u}(b) - \widehat{u}(d)
\end{eqnarray}
Finally, \eqref{prp:attract-eq14}, \eqref{prp:attract-eq15} and convexity
of $\widehat{u}(\cdot)\mapsto\widehat{u}(\cdot)^p$ jointly lead to the same
contradiction as above. This completes the proof that \eqref{rel-attr1} holds
under the second postulate as well.

We now show that \eqref{rel-attr2} holds under either of the postulated conditions. 
That is, we verify that $\widehat{u}(a)-\widehat{u}(b)
< \widehat{u}(c)-\widehat{u}(d)$ $\Leftrightarrow$ 
$\widehat{u}(a)^p-\widehat{u}(b)^p<\widehat{u}(c)^p-\widehat{u}(d)^p$.
Suppose first that $\widehat{u}(a)+\widehat{u}(b)=\widehat{u}(c)+\widehat{u}(d)$.
Let $\widehat{u}(a)-\widehat{u}(b) < \widehat{u}(c)-\widehat{u}(d)$ 
be true and assume to the contrary that 
\begin{eqnarray}
	\label{prp:attract-10} \widehat{u}(a)^p-\widehat{u}(b)^p & \geq & 
	\widehat{u}(c)^p-\widehat{u}(d)^p
\end{eqnarray} The former two assumptions imply
$\widehat{u}(a)<\widehat{u}(c)$, $\widehat{u}(b)>\widehat{u}(d)$ and
therefore
\begin{eqnarray}
	\label{prp:attract-11} \widehat{u}(c)>\widehat{u}(a)>\widehat{u}(b)
	>\widehat{u}(d)
\end{eqnarray}
Using again the convexity argument that revolved around 
\eqref{prp:attract-eq6} and \eqref{prp:attract-eq7} we get 
\begin{eqnarray}
	\label{prp:attract-12} \widehat{u}(a)^p + \widehat{u}(b)^p 
	> \widehat{u}(c)^p + \widehat{u}(d)^p
\end{eqnarray}
By \eqref{prp:attract-10} and \eqref{prp:attract-12} we now obtain
$\widehat{u}(a)>\widehat{u}(c)$, which is a contradiction. 
Conversely, suppose $\widehat{u}(a)^p-\widehat{u}(b)^p<\widehat{u}(c)^p-\widehat{u}(d)^p$
and assume to the contrary that $\widehat{u}(a)-\widehat{u}(b)
\geq \widehat{u}(c)-\widehat{u}(d)$. 
This and $\widehat{u}(a)+\widehat{u}(b)
= \widehat{u}(c)+\widehat{u}(d)$ jointly imply $\widehat{u}(a)>\widehat{u}(c)$
and $\widehat{u}(b)<\widehat{u}(d)$. 
Thus, we have $\widehat{u}(a)>\widehat{u}(c)>\widehat{u}(d)>\widehat{u}(b)$.
Using the above convexity argument once again we obtain
$\widehat{u}(c)^p+\widehat{u}(d)^p<\widehat{u}(a)^p+\widehat{u}(b)^p$.
Subtracting $\widehat{u}(a)^p-\widehat{u}(b)^p<\widehat{u}(c)^p-\widehat{u}(d)^p$
from this inequality yields $\widehat{u}(b)>\widehat{u}(d)$, a contradiction.

Finally, we establish \eqref{rel-attr2} under the postulate 
\begin{eqnarray}
	\label{prp:attract-eq16}
	\widehat{u}(a)^p+\widehat{u}(b)^p
	& = & \widehat{u}(c)^p+\widehat{u}(d)^p
\end{eqnarray}
Let 
\begin{eqnarray}
	\label{prp:attact-eq17}
	\widehat{u}(a)-\widehat{u}(b) 
	& < & \widehat{u}(c)-\widehat{u}(d),
\end{eqnarray}
and again assume to the contrary that \eqref{prp:attract-10} is true.
By \eqref{prp:attract-10} + \eqref{prp:attract-eq13} we get $\widehat{u}(a)
\geq \widehat{u}(c)$. This and \eqref{prp:attact-eq17} implies $\widehat{u}(b)
> \widehat{u}(d)$. But $\widehat{u}(a)\geq \widehat{u}(c)$ and \eqref{prp:attract-eq16}
also implies $\widehat{u}(b) \leq \widehat{u}(d)$. This is impossible.
Conversely, suppose $\widehat{u}(a)^p-\widehat{u}(b)^p<\widehat{u}(c)^p-\widehat{u}(d)^p$.
This and the postulated $\widehat{u}(a)^p+\widehat{u}(b)^p
= \widehat{u}(c)^p+\widehat{u}(d)^p$ jointly imply 
$\widehat{u}(c)>\widehat{u}(a)$ and $\widehat{u}(b)<\widehat{u}(d)$.
Together with the without-loss initial assumption whereby 
$\widehat{u}(a)>\widehat{u}(b)$ and $\widehat{u}(c)>\widehat{u}(d)$, 
this in turn implies $\widehat{u}(c)>\widehat{u}(a)>\widehat{u}(d)>\widehat{u}(b)$.
Assume to the contrary that $\widehat{u}(a)-\widehat{u}(b)
\geq \widehat{u}(c)-\widehat{u}(d)$. This is equivalent to
$\widehat{u}(d)-\widehat{u}(b)\geq \widehat{u}(c)-\widehat{u}(a)>0$.
Rearranging \eqref{prp:attract-eq16}, we also have 
$\widehat{u}(b)^p-\widehat{u}(d)^p = \widehat{u}(c)^p-\widehat{u}(a)^p$.
Since $\widehat{u}(\cdot)\mapsto\widehat{u}(\cdot)^p$ is a 
strictly increasing function, it follows from the above that 
the left hand side of this equation is negative while 
the right hand positive. This is a contradiction. Thus,
\eqref{rel-attr2} holds in this case too. \sqbox\\

\noindent \textit{\textbf{Proof of Proposition \ref{competition}.}}

Firm $i=1,2$ maximizes $\pi_i$ with respect to $q_i$ and $p_i$
taking the choices of the other firm $j\neq i$ as given.
Differentiating $\pi_i$ with respect to $p_i$, $q_i$ and simplifying we get
\begin{eqnarray*}
	\frac{\partial \pi_i}{\partial p_i} & = & 
	\dfrac{\left(\dfrac{p_j q_i}{(p_j q_i + p_i q_j}\right)^s 
		(p_j q_i + q_j (p_i - p_i s + q_i s))}{p_j q_i + p_i q_j}, \\
	\frac{\partial \pi_i}{\partial q_i} & = & 
	-\dfrac{\left(\dfrac{p_j q_i}{p_j q_i + p_i q_j}\right)^{(1 + s)}
	(p_j q_i^2 + p_i q_j (q_i - p_i s + q_i s))}{p_j q_i^2}
\end{eqnarray*}
Setting the two equations equal to zero yields the first-order conditions
\begin{eqnarray}
	\label{FOC-2} p_i^{*} & = & 
	\dfrac{q_i (p_j + q_j s)}{q_j (s-1)},\\
	\label{FOC-3} q_i^{*} & = & 
	\dfrac{p_i \sqrt{q_j} 
	\sqrt{q_j + 4 p_j s + 2 q_j s + q_j s^2}-p_i q_j - p_i q_j s}
	{2 p_j}
\end{eqnarray}
It can be checked upon rearranging these conditions 
in $\frac{q_i}{p_i}$ form (which, in particular, 
is a non-negative term) and simplifying 
that they cannot be satisfied simultaneously
under the assumption that $p_i,q_i,s\geq 0$ and $I>0$. 
This implies that there is no equilibrium 
where firms choose interior strategies. 
Since $q_i^{*}\leq p_i^{*}$ must hold, this fact and \eqref{FOC-2}, 
\eqref{FOC-3} together imply either $p_i^{*}=0$ or $p_i^{*}=I$. 
Because the latter (former) case is associated with a strictly 
positive (zero) profit, it follows that 
\begin{eqnarray*}
	p_i^* & = & I
\end{eqnarray*}
for $i=1,2$. Since the problem is symmetric, 
by \eqref{FOC-3} and $p_i^{*}=I$ we get
\begin{eqnarray*}
	q_i^{*} & = & 
	\frac{1}{2}\left(\sqrt{q_j (4 s I + q_j (1 + s)^2)}- q_j (1 + s)\right)
\end{eqnarray*}
for $i=1,2$.  Solving this system yields 
\begin{eqnarray*}
	q_i^{*} & = & \frac{sI}{2+s},
\end{eqnarray*} 
as claimed. 
The remaining assertions are verifiable 
by simple substitution.\sqbox\\

{
	
	\linespread{1}
	
	\printbibliography[notkeyword=software,heading=bibliography,title={Core References}]
	\printbibliography[keyword=software,heading=bibliography,title={Software References}]
	
}

\appendix

\section{Maximum-Likelihood Estimation in the Power Logit}

We extend the quadratic-logit analysis of Section 4.1 and clarify the properties 
and estimation aspects pertaining to the discrete-choice version 
of the more general power-logit model,where
\begin{eqnarray}
	\label{dc-power-logit1}	
	\rho_n(a_i,A) 
	& = 
	& \frac{e^{p\beta\cdot x_{ni}}}{\left(\sum\limits_{j=1}^{k} e^{\beta\cdot x_{nj}}\right)^p}\\
	\label{dc-power-logit2}	
	\rho_n(o,A) 
	& = 
	& \frac{\left(\sum\limits_{j=1}^{k} e^{\beta\cdot x_{nj}}\right)^p - 
		\sum\limits_{j=1}^{k} e^{p\beta\cdot x_{nj}}}{
		\left(\sum\limits_{j=1}^{k} e^{\beta\cdot x_{nj}}\right)^p}
\end{eqnarray}
Following \textcite{mcfadden73} and the ensuing literature, 
we show how the vector $\beta$ and scalar $p>1$ in 
\eqref{dc-power-logit1}-\eqref{dc-power-logit2} 
can be estimated by minimizing the log-likelihood 
function that emerges from this model. 
To this end, let us write
$$
\begin{array}{lll}
	Pr_{ni}	& \equiv & Pr_{ni}(\beta) \\
	& := & \rho_n(a_i,A),\\ 
	& & i\leq k,\\
\end{array}
\qquad\quad \text{and} \qquad\quad
\begin{array}{lll}
	Pr_{no}& \equiv & Pr_{no}(\beta)\\
	& := & \rho_n(o,A)\\ 
	& =  & 1-\sum_{i=1}^kPr_{ni}\\
	& > & 0.
\end{array}
$$
Next, let us denote by $y_n$ the $n$-th individual's 
observed decision at menu $A$. It is critical  
to distinguish between this decision being an active 
choice or choice of the outside option. 
To this end, we define the binary variables 
$y_{ni}$, $i=1,\ldots, k$, and $y_{no}$ by
$$y_{ni}:=
\left\{
\begin{array}{ll}
	1, & \text{if } y_n=a_i\\
	0, & \text{otherwise}
\end{array}
\right.
\qquad \text{and} \qquad
y_{no}:=
\left\{
\begin{array}{ll}
	1, & \text{if } y_n\neq a_i \text{ for all } a_i \in A\\
	0, & \text{otherwise}
\end{array}
\right.
$$
to account for the former and latter cases, respectively.
With these in place, the multinomial density 
for a given active-choice or opt-out decision made by agent 
$n$ can now be written as 
$$g_n(\beta,p)=Pr_{no}^{y_{no}}\prod_{i=1}^kPr_{ni}^{y_{ni}}.$$ 
Assuming an exogenous sample and covariates $x_{ni}$ for 
every agent $n\leq N$ and alternative $i\leq k$, 
the likelihood function that results from the $N$ independent 
decisions is now given by

\noindent
\begin{eqnarray*}
	L(\beta,p) & = & \prod_{n=1}^N g_n(\beta)\\
	& = & \prod_{n=1}^N \left(Pr_{no}^{y_{no}}
	\prod_{i=1}^kPr_{ni}^{y_{ni}}\right)
\end{eqnarray*}

\noindent
This leads to the log-likelihood function

\begin{eqnarray*}
	\nonumber LL(\beta,p)  & = & 
	\sum_{n}\sum_iy_{ni}\ln Pr_{ni} 
	+ \sum_{n}y_{no}\ln Pr_{no} \\
	\nonumber & = & 
	\sum_{n}\sum_iy_{ni}\ln 
	\left(\frac{e^{\beta\cdot 
			x_{ni}}}{\sum\limits_j e^{\beta\cdot x_{nj}}}\right)^p
	+ \sum_{n}
	y_{no}\ln\left(\frac{\left(\sum\limits_j e^{\beta\cdot x_{nj}}\right)^p
		-\sum\limits_{j}e^{p\beta\cdot 
			x_{nj}}}{\left(\sum\limits_j e^{\beta\cdot x_{nj}}\right)^p}\right)
\end{eqnarray*}

\noindent Recalling that $\beta=(\beta^1,\ldots,\beta^m)$ and $x_{ni}=(x_{ni}^1,\ldots,x_{ni}^m)$, 
the first-order conditions of its maximization with respect to $p$ and $\beta$ are

\begin{eqnarray*}
	\frac{\partial LL(\beta,p)}{\partial p} & = & 
	\sum_{n}\sum_{i}y_{ni}\left[\beta\cdot x_{ni}-
	\ln\left(\sum_{j}e^{\beta\cdot x_{nj}}\right)\right]\\
	& - & \sum_{n}y_{no}\left[
	\dfrac{\sum\limits_{j}(\beta\cdot x_{nj})e^{\widehat{p}\beta\cdot x_{nj}} -
		\sum\limits_j e^{\widehat{p}\beta\cdot x_{nj}}\ln\left(\sum\limits_j 
		e^{\beta\cdot x_{nj}}\right)}{\left(\sum\limits_j 
		e^{\beta\cdot x_{nj}}\right)^{\widehat{p}}-
		\sum\limits_{j}e^{\widehat{p}\beta\cdot x_{nj}}}\right]\\
	& = & 0,\\
	\frac{\partial LL(\beta,p)}{\partial \beta^l} & = & 
	p\sum_{n}\sum_iy_{ni}x_{ni}^l-p\sum_{n}\sum_iy_{ni}x_{ni}^l
	\left(\frac{e^{\widehat{\beta}\cdot x_{ni}}}{\displaystyle\sum_j 
		e^{\widehat{\beta}\cdot x_{nj}}}\right)\\
	& - & p\sum_{n}\sum_{i}y_{no}x_{ni}^l\left(
	\frac{e^{\widehat{\beta}\cdot x_{ni}}}{\displaystyle\sum_j 
		e^{\widehat{\beta}\cdot x_{nj}}}\right)	\\
	& + & p\sum_{n}y_{no} 
	\left[
	\dfrac{\left(\sum\limits_{j}e^{\widehat{\beta}\cdot x_{nj}}\right)^{p-1}
		\left(\sum\limits_{j}x_{nj}^le^{\widehat{\beta}\cdot x_{nj}}\right)
		-\sum\limits_{j}x_{nj}^le^{p\widehat{\beta}\cdot x_{nj}}}{
		\left(\sum\limits_j e^{\widehat{\beta}\cdot x_{nj}}\right)^p
		-\sum\limits_{j}e^{p\widehat{\beta}\cdot 
			x_{nj}}}
	\right]\\
	& = & 0,\qquad l = 1,\ldots,m
\end{eqnarray*}

\noindent 
Observing that $\sum_{i}y_{ni}+y_{no}=1$ holds by construction
and that $p>0$ enters all their terms multiplicatively,
the latter $m$ first-order conditions simplify to
\begin{eqnarray*}
	\sum_{n}\sum_iy_{ni}x_{ni}^l 
	& = & \sum_{n}\sum_ix_{ni}^l
	\left(\frac{e^{\widehat{\beta}\cdot x_{ni}}}{\displaystyle\sum_j 
		e^{\widehat{\beta}\cdot x_{nj}}}\right)\\
	& - & \sum_{n}y_{no} 
	\left[
	\dfrac{\left(\sum\limits_{j}e^{\widehat{\beta}\cdot x_{nj}}\right)^{p-1}
		\left(\sum\limits_{j}x_{nj}^le^{\widehat{\beta}\cdot x_{nj}}\right)
		-\sum\limits_{j}x_{nj}^le^{p\widehat{\beta}\cdot x_{nj}}}{
		\left(\sum\limits_j e^{\widehat{\beta}\cdot x_{nj}}\right)^p
		-\sum\limits_{j}e^{p\widehat{\beta}\cdot 
			x_{nj}}}
	\right]
\end{eqnarray*}
Thus, unlike the standard logit where the term 
appearing with a negative sign in the last equation is absent 
and where, by construction, the estimated 
$\widehat{\beta}$ ensures that empirical and average predicted 
frequencies of active-choice alternatives coincide 
\parencite{train09,greene-hensher}, 
the presence of the said term here clarifies that this is no longer true 
in the power logit when deferral choices are present in the data.

\end{document}